   \documentclass[sigconf]{acmart}
\usepackage{tabularray}
\usepackage[graphicx]{realboxes}
\usepackage{fontawesome}

\usepackage{pifont} 
\usepackage{pgfplotstable} 
\usepackage{paralist}
\usepackage{amsmath}
\usepackage{multirow}
\usepackage{shortcuts}

\usepackage{epsfig,endnotes}
\usepackage{tikz}
\usepackage{pgfplots}
\usepackage{subcaption}
\usepackage[ruled,vlined,linesnumbered]{algorithm2e}
\usepackage[toc,page]{appendix}
\usepackage{amsfonts}
\usepackage{url}
\usepackage{array}
\usepackage{booktabs,adjustbox}
\usepackage{etex}
\usepackage{color}
\usepackage{pstricks}
\usepackage{wasysym}
\usepackage{shortcuts}
\usepackage{paralist}
\renewcommand{\paragraph}[1]{\vspace{0.05in}\noindent{\textbf{#1}.}}
\usepackage[justification=justified]{caption}
\usepackage{pgfplots}
\usetikzlibrary{arrows}
\usetikzlibrary{patterns}
\usepackage{listings}
\usepackage{multicol}
\usepackage{lipsum}
\usepackage{adjustbox}
\usepackage[font=bf,skip=\baselineskip]{caption}
\usepackage{tabularx}
\usepackage{multirow}

\usepackage[english]{babel}
\usepackage[utf8x]{inputenc}
\usepackage{listings}
\usepackage{color}
\usepackage{subcaption}
\usepackage{latexdiff}
\let\labelindent\relax

\usepackage{enumitem}
\usepackage{graphicx}

\setitemize[1]{itemsep=0pt,partopsep=0pt,parsep=\parskip,topsep=5pt}
 
\usepackage{diagbox}

\usepackage{hyperref}
\usepackage{cleveref}
\hypersetup{
    colorlinks = true,
    linkcolor = blue,
    anchorcolor = blue,
    citecolor = blue,
    filecolor = blue,
    urlcolor = blue
}

\usepackage{amsmath} 
 
\newif\ifdiffvar
\diffvarfalse

\usepackage[T1]{fontenc}
\usepackage{dashrule}

\newcommand{\tickcoloredYes}{{\textcolor{green}{\checkmark}}}
\newcommand{\tickcoloredNo}{\textcolor{red}{\ding{55}}}
\newcommand{\tickYes}{\checkmark}

\DeclareTextFontCommand{\breakabletexttt}{\ttfamily\hyphenchar\font=45\relax}

\newcommand{\wechat}{{\sc WeChat}\xspace}

\newcommand{\tool}{{\sc Tools}\xspace}
\newcommand{\totalnumber}{{\sc $8{,}824$}\xspace}

\usepackage[T1]{fontenc}
\usepackage{pmboxdraw}
\usepackage{xcolor}
\def\SFii{\textcolor{blue}{\textSFii\textSFx}}
\def\SFviii{\textcolor{blue}{\textSFviii\textSFx}}

 \usepackage{tikz}

 \renewcommand\footnotetextcopyrightpermission[1]{} % removes footnote with conference information in first column
\newcommand{\AY}[1]{{\an{AY}{#1}}}
\newcommand{\ZY}[1]{{\an{ZY}{#1}}}

\newcommand{\token}{{\texttt{LT}}\xspace}
\newcommand{\appsecret}{{\sc \texttt{MK}}\xspace}
\newcommand{\appsecrets}{{\sc \texttt{MKs}}\xspace}
\newcommand{\sessionkey}{{\sc \texttt{EK}}\xspace}

\crefname{figure}{Figure}{Figures}
\crefname{equation}{Equation}{Equations}
\crefname{table}{Table}{Tables}
\crefformat{section}{#2\S#1#3}
\crefname{listing}{Listing}{Listings}
\crefname{algorithm}{Algorithm}{Algorithms}
\crefname{algocf}{Algorithm}{Algorithms}
\crefname{appendix}{Appendix}{Appendices}
 \usepackage[all]{nowidow}
 \setcopyright{none} % No copyright notice required for submissions
\acmConference[This is a preprint of our CCS 2023]{ACM Conference on Computer and Communications Security} 
\acmYear{2023}
\usepackage{fancyheadings}
  \usepackage{fancyhdr}
 
\begin{document}

\title{Don't Leak Your Keys: Understanding and Exploiting the Cryptographic Access Control in  Mini-Programs}

\title{Don't Leak Your Keys: Understanding, Measuring, and Exploiting the AppSecret Leaks in Mini-Programs}

\author{Yue Zhang}
\affiliation{%
  \institution{The Ohio State University}
%   \city{Columbus}
%   \state{OH}
  %\country{USA}
}
\email{zhang.12047@osu.edu}

\author{Yuqing Yang}
\affiliation{%
  \institution{The Ohio State University}
%   \city{Columbus}
%   \state{OH}
  %\country{USA}
}
\email{yang.5656@osu.edu}

\author{Zhiqiang Lin}
\affiliation{%
  \institution{The Ohio State University}
%   \city{Columbus}
%   \state{OH}
  %\country{USA}
}
\email{zlin@cse.ohio-state.edu}
%\title{I Don't Trust Front-ends Anymore:\\ Understanding and Measuring the Misuse of Cryptographic Keys in Mini-Programs}
%\title{Cryptographic Access Control in Mini-Programs:\\ A Closer Look at Key Leakages}
%\author{\#33}

\definecolor{codegreen}{rgb}{0,0.6,0}
\definecolor{codegray}{rgb}{0.5,0.5,0.5}
\definecolor{codepurple}{rgb}{0.58,0,0.82}
\definecolor{backcolour}{rgb}{1,1,1}
\lstset{language=[AspectJ]Java,
  backgroundcolor=\color{backcolour},
    commentstyle=\color{codegreen},
    keywordstyle=\color{magenta},
    numberstyle=\tiny\color{codegray},
    stringstyle=\color{codepurple},
    basicstyle=\footnotesize,
    linewidth=0.45\textwidth,
    breakatwhitespace=true,
    frame = single,
    breaklines=true,
    captionpos=b,
    keepspaces=true,
    numbers=left,
    numbersep=6pt,
    showspaces=false,
    showstringspaces=false,
    showtabs=false,
    tabsize=2,
     morekeywords={tt,success,fail,options}
    }

 \begin{abstract}
\ignore{During the past few years, we have witnessed an explosion of mobile mini-programs (in the same scale as apps in Google Play), which are small programs running inside privacy-rich super apps such as \wechat.  
To protect the user-sensitive data (e.g., phone numbers, daily walked steps) of mini-programs from being leaked, cryptography-based access control mechanisms have been implemented, in which the sensitive data is encrypted in the front-end and decrypted in the back-end.  
Consequently, the key to encrypt and decrypt the sensitive data needs to be properly managed. Unfortunately, we show in this paper that there are \leakedapps mini-programs that have leaked the corresponding keys by hard-coding them in the front-end of mini-programs. More importantly, we also show that such key leakage can allow attackers to 
launch various novel and unique attacks including account hijacking, promotion abuse, and service theft.  
We have disclosed the cryptographic key leakage vulnerabilities  as well as the list of the vulnerable mini-programs to  {\it Tencent}, who awarded us with bug bounties and also recently released a new API to defend against the attacks based on our recommended defense.  
}

%In recent years, the use of mobile ``mini-programs'' --- small programs that run within privacy-rich super apps like \wechat --- has exploded in popularity. To protect sensitive user data such as phone numbers and daily walked steps from being leaked, these mini-programs use cryptography-based access control mechanisms that encrypt sensitive data in the front-end and decrypt it in the back-end. Consequently, the key to encrypt and decrypt the sensitive data needs to be properly managed. Unfortunately, in this study we have found that \leakedapps mini-programs have leaked the keys used for this encryption by hard-coding them in the front-end of the mini-programs. This key leakage leaves mini-program users vulnerable to a range of attacks, including account hijacking, promotion abuse, and service theft. We have reported these vulnerabilities and the list of vulnerable mini-programs to Tencent, who awarded us with bug bounties and also recently released a new API to defend against these attacks based on our recommendations.

Mobile mini-programs in \wechat have gained significant popularity since their debut in 2017, reaching a scale similar to that of Android apps in the Play Store. Like Google, Tencent, the provider of WeChat, offers APIs to support the development of mini-programs and also maintains a mini-program market within the \wechat app. However, mini-program APIs often manage sensitive user data within the social network platform, both on the \wechat client app and in the cloud. As a result, cryptographic protocols have been implemented to secure data access. In this paper, we demonstrate that \wechat should have required the use of the ``appsecret'' master key, which is used to authenticate a mini-program, to be used only in the mini-program back-end. If this key is leaked in the front-end of the mini-programs, it can lead to catastrophic attacks on both mini-program developers and users. Using a mini-program crawler and a master key leakage inspector, we measured 3,450,586 crawled mini-programs and found that \leakedapps of them had leaked their master keys, allowing attackers to carry out various attacks such as account hijacking, promotion abuse, and service theft. Similar issues were confirmed through testing and measuring of Baidu mini-programs too. We have reported these vulnerabilities and the list of vulnerable mini-programs to Tencent and Baidu, which awarded us with bug bounties, and also Tencent recently released a new API to defend against these attacks based on our findings. %ZQ{people may challenge "are they really based on our recommendations?"}
    \end{abstract}
 \maketitle

 \section{Introduction}
   \vspace{2mm}
    \label{sec:intro}

A mini-program, which is a small program restricted in size, is gaining popularity on social network platforms such as \wechat, which is the third most popular messenger app with 1.2 billion monthly active users~\cite{Mostpop46:online}. These mini-programs have greatly expanded the capabilities of super-apps, such as \wechat, which has almost become an ``operating system'' offering more than 900 APIs~\cite{wechatapi} for executing more than 4 million mini-programs~\cite{wechat3M}. These mini-programs cater to various daily needs of users, including online shopping in virtual stores and scan-to-buy in physical stores. Among the popular mini-programs running on the \wechat platform,  \texttt{PinDuoDuo}, a group-buying app, is such an example.

While many social network platforms, like Facebook~\cite{facebookAPIs}, have provided APIs for third-party developers to access and use the social network information collected, super apps such as \wechat have taken it further and made it even more open with their mini-program paradigm. Mini-programs can access more generalized user-specific data directly in the platform, maintained by both the local host-app and remote cloud, through uniformed APIs. For instance, by invoking \texttt{getPhoneNumber}, a \wechat mini-program can retrieve a user's telephone number. A telephone number in mobile super apps is a crucial identifier used to bind the mini-program account to a specific user, and the super apps even allow mini-programs to authenticate their users solely through the phone number without any passwords. Furthermore, in addition to local resources and user-specific data, \wechat provides online paid services like Optical Character Recognition (OCR), which converts images of text to electronic text, to enrich the functionalities of the mini-programs without adding heavy barriers. \looseness=-1

Super apps have implemented access control mechanisms and cryptography protocols to secure services access and data transmission to protect privacy-sensitive data and paid services from potential data leaks and abuse by mini-programs. For instance, when accessing sensitive user records stored in \wechat servers, \wechat encrypts the data with a user-specific encryption key (\texttt{EK} or {\tt session key} in Tencent's terminology), which the mini-program back-end requests with the mini-program's master key (\texttt{MK} or {\tt AppSecret}), the mini-program ID, and the user's login token (\texttt{LT}). The mini-program back-end then decrypts the data on the back-end using the retrieved \texttt{EK}. Similarly, when accessing paid services provided by \wechat, the mini-program back-end requests an access token with \texttt{MK}, mini-program ID, and uses the obtained access token to invoke the services. However, not all developers understand the implications of the untrustworthiness of the mini-program front-end, and some mistakenly distribute \texttt{MK}s to the front-end of the mini-program, as confirmed with our manual analysis of several well-known mini-programs that store their \texttt{MK}s in the mini-program code, allowing an attacker to easily obtain their \texttt{MK}s. \looseness=-1

This study systematically examines \appsecret leaks in \wechat to understand their prevalence, root causes, and consequences. Our findings can be applied to other super apps, given the similar mechanisms they share. In particular, we show that \appsecret leaks can result in severe consequences, including user account hijacking, promotion abuse, and service theft. For example, in \wechat, a common practice to index users is through their phone numbers. With the leaked {\tt MK} of a vulnerable mini-program, an attacker can query the \sessionkey from the \wechat servers to decrypt any encrypted data delivered from the servers. This enables access to all data maintained by the victim's account, such as the user's citizen ID and shipping address. Additionally, the attacker can cause financial losses to the victim, such as gaining cashback and coupons from vendors by manipulating sensitive data, such as group chat information. Finally, with the obtained \texttt{MK}, the attacker can consume paid services purchased by the victim for free.

To gauge the severity and prevalence of the \appsecret leak among mini-programs, we have conducted a measurement study in this paper. Specifically, by scanning a dataset of \totalanalyzedapps mini-programs, we discover \leakedapps mini-programs that had leaked their \appsecret{s}. Our evaluation also reveals that vulnerable mini-programs are not limited to unknown developers, but include popular mini-programs from high-profile vendors such as \textit{Nestle}, \textit{HP}, and \textit{Tencent}. We have reported the \appsecret leakage vulnerabilities and the list of vulnerable mini-programs to {\it Tencent} and been awarded with bug bounties. Currently, {\it Tencent} is actively working with developers to fix this vulnerability, and some mini-programs have already removed \texttt{MK}s from their code. Furthermore, {\it Tencent} recently released a new API to defend against attacks based on our findings~\cite{checkEncryptedData}. \looseness=-1

\ignore{
In this paper, we seek to understand the severity and prevalence of \appsecret leakage attacks, and meanwhile develop tools to detect the vulnerable mini-programs in the \wechat eco-system. To this end, we 
To measure the impact and severity of our attack, we implemented a tool and tested it with more than 1.2 million mini-programs.  our study revealed a concerning situation. We found that 3\% of the mini-programs leaked \appsecret, over 2\% of these mini-programs contained potential \sessionkey leakage. \ZY{Allen, please add some interesting results here to impress our readers}
}

\ignore{
That is, instead of enforcing those mission critical security measures directly, \wechat gives the power of implementing security functionalities to developers, and those security functionalities  may server as safe-guard of sensitive data maintained by \wechat. This may causes grave consequences, as a sloppy third-party developers may implement the security measures in various faulty ways. For example, we observed that many mini-programs mistakenly hard-coded the aforementioned app-secrets in their source codes and as a result, the attackers can arbitrarily read, modify, and relay the sensitive information to perform an Man-In-The-Middle (MITM) attack. %In another of our example, \wechat provide security levels for Bluetooth pairing, 
We denoted this type of attack as \textsf{Coffee} (\textbf{C}ost \textbf{OF} \textbf{F}r\textbf{EE}dom) attack. 

The \textsf{Coffee} attacks not only occur in these ``Add-on'' security functionalities that secure communication between the mini-program front-end and its back-end, but also occur in some other security related components. In the Bluetooth communication, instead of enforcing ``build-in'' secure configuration, \wechat allows the developers to customize the (or even disable) the secure configurations,  which allows the attacker to achieve unauthorized data access or MITM attacks.  We also found that in the cross-app communication, \wechat have removed a build-in whitelist mechanism that allowed the mini-programs to communicate with only ``trusted'' mini-programs, to give the flexibility to the developers.    With such a build-in white-list removed, the communication channel between two apps should be secured by the ``Add-on'' access control,  which implemented by the developers. 
However, no instructions are given by \wechat by the time of our writing,  and as a result, the developers may ignore implementing the access control, leaving their cross-mini-program channels unprotected. 

To understand the impact of \textsf{Coffee} attack, we design an measurement tool, \tool, and analyzed \totalnumber mini-programs. We found that .....

The most important lesson learnt from our research is that it should never to assume that all developers are benign and hence offer the developers the freedom of security mechanism implementation, as the developers may not have the patience to read the documentations and understand the mechanism and significance of properly implementing secure measures well due to the frequent document updating and various other reasons. To ensure the security of such a complex paradigm that involves not only various platforms and devices but also heterogeneous back-ends, the designer of the platform are supposed to be aware that the mini-program developers should never be trusted. Therefore, instead of letting the mini-program developers to implement security features on top of APIs provided, the framework should have encapsulated all the vital security mechanisms well, and make them transparent to the developers.
}

\paragraph{Contributions} We make the following contributions: %The contributions are outlined as follows:
\begin{itemize}
    \item {\bf Systematic Understanding (\S\ref{sec:understanding}).} We are the first to systematically examine the sensitive resource access protocols of mini-programs, resulting in the discovery of \appsecret leakage vulnerabilities across multiple platforms such as WeChat.

  \item {\bf Empirical Measurement (\S\ref{sec:measurement}).} We develop a measurement tool and evaluate it with a large set of \totalanalyzedapps mini-programs. The result shows that currently around \leakedapps of mini-programs that contain \appsecret leakage vulnerability.
    
    \item {\bf Practical Attacks (\S\ref{subsec:attackworkflow}).} We demonstrate two types of novel attacks with leaked \appsecrets: attacks against sensitive resources and attacks against cloud services.  
    %
     %
    %\item {\bf Practical Tool.} We develop a practical pipe-lined analysis framework that is able to automatically crawl mini-programs using keywords and detect the mini-programs that are subject to the \appsecret leakage attacks. Our crawler is developed using mini-program APIs and is more efficient and practical than existing crawlers. %The source code will be made public available.
    %
   % \item {\bf Real-world Implications.} 
    %Our evaluation with the automatically crawled \totalanalyzedapps mini-programs, the largest set of mini-programs,  shows that \leakedapps mini-programs that contain \appsecret leakage vulnerability. 
  %  We perform case studies on some vulnerable mini-programs from high-profile vendors such as Nestle, Bank of China and even Tencent themselves. 
  We show that these attacks can have devastating impacts in the super app ecosystem, such as hijacking user's account,  manipulating user's sensitive data, or consuming cloud services for free. %\textcolor{red}{We found that there are \leakedapps that are vulnerable to those attacks. }
    
    %attack case studies further show that with the leaked \appsecret, the attacker can log into arbitrary user's account, and manipulate various sensitive data for fun and profit (e.g., earning cashback and coupons).   

   \item {\bf Possible Countermeasures (\S\ref{sec:discussion}).}  In addition to the responsible disclosure of this vulnerability as well as the list of \leakedapps mini-programs that leaked {\tt MK}s, we also shed light on the possible mitigations and preventions, particularly on how super app vendors could have fixed this issue.
    %\item We made responsible disclosure to \Tencent and our findings have been acknowledge by \Tencent developer team. \AY{Besides, we will release our analysis tool for the community to call for the awareness of potential vulnerabilities in this emerging paradigm.}

\end{itemize}

%\paragraph{Roadmap} 

%Such a practice 

%usually the weakest link of a system,     

% \ZY{Need to explain why the \sessionkey or app secret is needed} 
%The benefit of using such a cryptographic key instead of using an on-the-shelf interface is that  That is, instead of enforcing those security measures directly, \wechat provided more flexible interfaces for developers to customize, allowing the developers to have a balance between security and other considerations (e.g., usability).

\ignore{

It is surprising that such a mini-program platform involved with huge user base and massive sensitive data would allow developers to maintain the most vital keys on their own. To ensure the security of such a complex paradigm that involves not only various platforms and devices but also heterogeneous back-ends, the designer of the platform are supposed to be aware that the mini-program developers should never be trusted. Therefore, instead of letting the mini-program developers to implement security features on top of APIs provided, the framework should have encapsulated all the vital security mechanisms well, and make them transparent to the developers. Unfortunately, we found that for some of their most important security measures like maintaining \sessionkey, wechat merely provides a set of APIs and descriptions of how to properly maintain sensitive data in their documentations, rather than a mechanism to check whether the security mechanisms are properly implemented.

It is never practical or secure to assume that all developers are benign and  hence offer the developers the freedom of security mechanism implementation. On one hand, the developers may not have the patience to read the documentations and understand the mechanism and significance of properly implementing secure measures well. On the other hand, the mini-program community is still actively introducing new features and frequently updating APIs and documentations, which makes it impossible for every developer to keep the most up-to-date understanding of the entire mini-program framework. Therefore, without a set of properly designed APIs that encapsulates necessary security mechanisms, vulnerability is inevitable when developers make mistakes when implementing these mechanisms on their own.

% Nevertheless, the developers are not the ones to be blamed. By looking into the official documentations, {\red{we found that although wechat had realized the challenge of enhancing the security of mini-program platforms and introduced mechanisms including scope management and cloud functions, their security countermeasures unfortunately merely covers minimal security requirements such as permission-like management and basic encryptions.}}\ZY{The description itself doesn't have too much connections with the previously provided example. In that example, you talk about \sessionkey and back-end etc. In the descriptions here, which based on my understanding, should be an explanations for the example,  however, you talk about some other stuffs including the scope management and cloud functions. I don't think the logic here flows.    }
% unfortunately do not cover all scenes due to the complexity of the paradigm. 
As to additional security measures, \wechat framework merely provided instructions and APIs for developers themselves to implement part of the security measures at their own risks, instead of providing a uniformed encapsulation in the SDKs or directly making these mechanisms transparent to developers.

This brings a common dilemma faced by app-in-app paradigms that combine native apps and webapps together in a sandbox-like subsystem: the dilemma between enhancing security and supporting versatile functionality. When the mini-program framework involves complex and heterogeneous platforms, devices, and self-configured back-ends, the framework can no longer enforce all security measures by itself with built-in mechanisms, because it can no longer control every involved entities.  \ZY{You story totally changed here. Our argument is being a cross-platform framework, \wechat should take care of everything. Now, you are saying that the framework can no longer enforce all the security measures by itself. So, what is our insight? The point is that \wechat can do something that makes the system more secure but they didn't do, not \wechat cannot do the security measures, so they have to leave it for the developers. In the second case, \wechat should not be blamed. } As a result, the framework may choose to prioritize their functionality requirements, offering freedom to mini-program developers at the expense of failure in splitting built-in and add-on security mechanisms properly to minimize the security impact of misuse and misconfiguration of sloppy developers.

In this paper, we investigate the communication security problems lying underneath the prosperity of mini-program freedom with static analysis of mini-program packages. Our study reveals a concerning situation. We discovered that foo and bar. \AY{I think this is going to be the first draft}

}

\section{Background}
 \label{sec:back}

\ignore{
\begin{itemize}
\item \textbf{Social App Vendor.} Through the mini-program paradigm, social network app providers can free their hands from constantly keeping designing new features. Instead,  they can focus more on the core building blocks and functionalities (e.g., the APIs and the security mechanisms of the ecosystem). \AY{Moreover, the highly encapsulated and easy-to-use paradigm lowers the complexity and cost of app development, thus attracting  more vendors, even non-IT vendors, which brings more stickiness of vendors and their users, as well as more profit.}%with the functionalities designing %without considering the user's preferences, and

% With the core building blocks, they can unleash the power and creativity from the numerous third-party developers and let them, as well as the market decide the best mini-programs satisfying customer's needs. Meanwhile, customers will also be satisfied with the social apps and cannot easily leave them.  \looseness=-1

%The mini-programs are designed to lighten the social network apps, and they have to use multiple approaches to save resources as much as possible. For example, their sizes are now limited to a few MBs (e.g., \wechat does not allow mini-programs that larger than 12 MB~\cite{sizelimit2021} to be uploaded onto their market), which is way small than the native apps. 
 
\item \textbf{Third Party Developers.} The third party developers can benefit tremendously from this novel paradigm: First, they can make use of the massive amount of user personal and social data from the social app
% the social network apps have collected a lot of user information, and now those information is open to those third party developers
(e.g., using \texttt{getPhoneNumber} to obtain the user's phone number), and they can even use these information to promote their services such as using sharing information to track users who recommended their mini-programs to other users and award the users. Second, many core functionalities of mini-programs can be accomplished conveniently with the mini-program APIs tailored to the need of mini-program era: for instance, the account creation in mini-programs is just one click away by letting users to link the social app account to the mini-program.

% the third party developers can now promote their products and services utilizing the user social networks via various approaches provided by the host app, e.g., via QR code sharing, hyperlink sharing, ``red-packet'' cashback, and even couponsonce hosted by the social network apps, the mini-programs developed by third-party developres for example, mini-programs can leverage multiple features (e.g., QR ocde sharing, hyperlink sharing, ``red-packet'' cashback, and even coupons) provided by the host apps to promote their products and penetrate the market.  %For example, the developers may give the users some cashback or coupons to let the user promote their products. \looseness=-1
 
\item \textbf{Users.} The users will experience less size increment and update frequency of the social network apps, whereas they have more freedom to customize the functionality by adding the mini-programs they need.
% bothered by the increasing size and frequent updates of the social network apps, given that they can customize the functionalities freely. 
Moreover, the strick size restriction and install-less feature (e.g., the users can use a mini-program by simply clicking the mini-program icon~\cite{mini-programS51:online}) offered by the mini-program can significantly enhance the user experiences. Finally, \AY{spcailly tailored mini-program functionalities such as the one-click log in feature makes the usage of mini-programs even more convenient in that the users no longer need to register multiple accounts, receive verification codes and remembering and entering the passwords when loging in.}
% given the third party developers can now reuse the data provided by the social network apps, the users do not need to register multiple accounts on different mini-programs. Instead, they can simply use the data collected (e.g., phone number) by the social network apps to login any mini-programs, which now is supported by the multiple mini-program providers. 

\end{itemize}

}

%As such, the mini-programs have multiple novel fancy evaluations when compared with the traditional paradigms. 

%This creates a win-win situation, the user can 

%\subsection{Sensitive Resource Access}
%\label{subsec:data}

%With the convenient and secure access provided by social network apps in the mini-program paradigm, mini-programs can access various resources, including system resources (e.g., Bluetooth, GPS data), user specific sensitive data (e.g., phone number, user information), and online services (e.g., OCR services).   
%from both the local devices of mini-program users via social network app and the cloud services from social network app server or the developer's back-end server. 

%Among these resources, we mainly focus on the resources secured with MK-centric resource access protocol given its novelty, since other resources such as system resources are secured by the classical in-app permission mechanism (which has been studied by prior works~~\cite{lu2020demystifying}). 
\subsection{Sensitive Resource Access by Mini-programs}
There are two types of resource that can only be accessed by mini-programs in \wechat: (1) sensitive data, which is user-specific and requires mini-program developers to fetch a decryption key using the \appsecret to access, and (2) cloud services, which are mini-program-specific and require mini-program developers to fetch an access token using the \appsecret to access. In the following, we discuss these two types of sensitive resource access in more detail.
\begin{itemize}
\item \textbf{Sensitive Data Access}.
Sensitive data refers to data generated by end-users and collected by mini-programs. For example, by examining the APIs provided by \wechat and their official documentation~\cite{wechatapi}, we have identified multiple privacy-sensitive data, as shown in \autoref{tab:wechatdata} (note that other super-apps such as Baidu have similar APIs that can be invoked by their mini-programs to access sensitive data). This data includes user information associated with a particular user, such as their nickname and phone number, which is kept on \wechat servers. When accessed by third-party mini-programs through APIs, \wechat will first encrypt the data and allow them to be decrypted only with a decryption key that is fetched from the \appsecret on the mini-program's back-end.
\looseness=-1

% Sensitive data refers to the type of data generated from end-users and collected by the mini-programs. For example,  having examined each API provided by \wechat and their official documentation~\cite{wechatapi}, we find that there are multiple privacy-sensitive data, as shown in \autoref{tab:wechatdata} (note that other super apps have the similar set of APIs that can be invoked by their mini-programs to consume the sensitive data).  For example,  there is user information associated with a particular user (\eg, the nick name,  telephone number). These sensitive information is kept at \wechat servers. When
% they are accessed by 3rd-party mini-programs through
% APIs, \wechat will encrypt them first and allow them to
% be only decrypted with a decryption key that is fetched from \appsecret at mini-program’s back-ends. \looseness=-1

\item \textbf{Cloud Service Access}. Cloud services are the services provided by \wechat to mini-program developers for freeing them from re-implementing complex functionalities such as Optical Character Recognition (OCR) services. Some services are not provided for free (e.g., fraud detection services provided by \wechat cost around \$5,000 per million invocations). Having investigated the services provided by \wechat on the online developer dashboard, we find that those services can be grouped into 3 categories as shown in \autoref{tab:wechatdata}:  (1) AI services (e.g., for AI chat bots);  (2) Security services,  which detects potential user risks as well as fraud risks; and (3) Map services (e.g., for  searching and resolving locations).  When they are accessed by 3rd-party mini-programs through APIs, \wechat requires a correct access token obtained from \appsecret at the mini-program’s back-ends.\looseness=-1
\end{itemize}

 \begin{table}[]
\scriptsize
 
 \setlength\tabcolsep{2pt}
 
    \resizebox{0.485\textwidth}{!}{ 
\begin{tabular}{llcc}\toprule
\textbf{Name} & \textbf{API} & \textbf{Encrypted?} & \textbf{Price} \\\midrule
\multicolumn{4}{c}{\textbf{Sensitive Data}} \\\midrule  
%  {\textbf{Sensitive}} &  {\textbf{API Name () /}} & &  \\
% \textbf{Data}  &  \textbf{Object Field Name}  &
%\\

%\textbf{User Information}    &   &   &   \\
Phone number       & \texttt{getPhoneNumber()}       &  \tickYes& - \\  
%\SFviii Language & \texttt{obj.lang} & \tickNo    &  \tickYes     \\
User Info & \texttt{wx.getUserProfile()} &   &  \\
~~~\SFii Gender      & \SFviii\texttt{info.gender}       &  \tickYes& -   \\
~~~\SFii Nick name    & \SFviii\texttt{info.nickName}  &  \tickYes& -    \\

~~~\SFii Avatar   & \SFii\texttt{info.avatarURL} &  \tickYes& -  \\  
 
%\textbf{Social Information}        &      &     &      \\  
  Shared Info        & \texttt{wx.getShareInfo()}     &  \tickYes& -   \\  
 Promoting Messages        & \texttt{wx.authPrivateMessage()}      &  \tickYes& -   \\ 
  Group Chat Info       & \texttt{wx.getGroupEnterInfo()}      &  \tickYes& -   \\  
%\midrule
%\textbf{Sports Information}        &      &     &      \\  
  WeRundata       & \texttt{wx.getWeRunData()}  &  \tickYes& -  \\ \toprule
\multicolumn{4}{c}{\textbf{Cloud Services}} \\\midrule 
 
\textbf{AI services}    &   &   &   \\
{OCR Services}    &   &   &   \\
%{Optical Character Recognition (OCR) Services}    &   &   &   \\
~\SFviii Bank account       & \texttt{openapi.ocr.bankCard}       &  \tickNo& $\sim$\$ 1,000 \\ 
~\SFviii Business License      & \texttt{openapi.ocr.businessLicense}       &  \tickNo& $\sim$\$ 1,000 \\ 
~\SFviii Drive License       & \texttt{openapi.ocr.driveLicense}       &  \tickNo& $\sim$\$ 1,000 \\ 
~\SFviii National ID       & \texttt{openapi.ocr.idCard}       &  \tickNo& $\sim$\$ 1,000 \\ 
~\SFviii Regular Text      & \texttt{openapi.ocr.plainText}       &  \tickNo& $\sim$\$ 1,000 \\ 
~\SFii License plate  &  \texttt{openapi.ocr.vehicleLicense} &  \tickNo& $\sim$\$ 1,000  \\
 AI Chat Bot      & \texttt{openapi.ans\_node\_name}       &  \tickNo& 0 \\ 
  AI Products Classification      & \texttt{goodclass2}       &  \tickNo& 0 \\ 
 Translation      & \texttt{multilingualMT}       &  \tickNo& 0\\ 
 Jokebot      & \texttt{jokebot}       &  \tickNo& 0\\ 
 Products Info Extraction      & \texttt{goodinfo}       &  \tickNo& 0 \\ \midrule
\textbf{Security Services}    &   &   &   \\
\SFviii Black Market Report      & \texttt{weixinSecintelligenceresp}       &  \tickNo& 0\\ 
\SFviii Fraud Detection     & \texttt{weOpensecRiskservice}       &  \tickNo&  $\sim$\$ 5,000 \\ 
\SFii User Risks Detection    & \texttt{weOpenSecuseracctRiskLevel}       &  \tickNo& $\sim$\$ 5,000 \\ \midrule
\textbf{Map Services}    &   &   &   \\
\SFviii Poi Search & \texttt{poisearch}       &  \tickNo& $\sim$\$ 260 \\ 
\SFviii Address Resolution    & \texttt{geoc}       &  \tickNo&  $\sim$\$ 260 \\ 
\SFviii Coords Conversion    & \texttt{coordTrans}       &  \tickNo&  $\sim$\$ 260 \\ 
\SFii Poi Suggestion    & \texttt{poiSuggestion}       &  \tickNo& $\sim$\$ 260 \\ 

\bottomrule
\end{tabular} 
 }
 
\caption{Summary of sensitive resources managed by \wechat. The price is for 1 million's invocations of the services~\cite{aiservices}. }%Note that E stands for Encryption, $P_{app}$ stands for Permission Check enforced by \wechat, $P_{os}$ stands for Permission Check enforced by OS. }%\ZQ{pls remove R column}}%, and R stands for Rate Limiting.}  
\label{tab:wechatdata}
 \end{table}
 
\subsection{Comparison of Mini-Programs, Mobile Apps, and Web Apps} 

While \wechat has almost become an ``operating system'' for mini-programs, it is different from traditional mobile operating systems and web browsers that also host web apps. Specifically, like web apps, mini-programs do not require installation, while mobile apps do. Super app platforms store users' data (e.g., phone numbers, billing addresses) on the cloud, allowing mini-programs to access this information via APIs, whereas mobile operating systems and web browsers only offer APIs for apps to consume locally stored data (e.g., location, photos, and videos). Super apps provide cloud APIs for mini-programs to utilize, whereas mobile operating systems and web browsers do not, relying instead on third-party online services (e.g., Amazon cloud services). The mini-program ecosystem natively supports encryption based access protocols (for accessing sensitive data), token-based access protocols (for using paid services), and key management protocols (for distributing cryptographic keys), while web and mobile apps can implement such protocols independently, but their platforms do not inherently provide them.
Finally, both mini-programs and mobile apps must be vetted by their respective platforms before users can access them, whereas web apps do not require vetting, as anyone can create web pages for others to access.

%Super apps' threat models assume that mini-program front-ends cannot be trusted, requiring sensitive data or services to be consumed at the back-end, an assumption that does not apply to mobile or web apps \ZQ{Mobile and web app should not also assume the trust of the front-ends. I feel your statement will raise red flags}. 
\ignore{

\subsection{Cryptographic Access Control}
\label{subsec:accesscontrol}

Cryptographic access control~\cite{cryptoacl} is widely used in computer security, and based on the types of keys, the access control can also be classified as asymmetric cryptographic access control and symmetric cryptographic access control:

\begin{packeditemize}
\item \textbf{Asymmetric Cryptography~\cite{simmons1979symmetric} Based Access Control}, also known as public key based access control, used a private-public key pair to control the access of resources. At a high level, it requires the private key to be securely maintained by the user who attempts to share the resources, and distribute the public key to others who intend to access those resources. In particular, when a user shares a resource (e.g., a message), she first produces a symmetric key and uses this key to encrypt it, and then encrypts the symmetric key using the receiver's public key. The cipher (including the encrypted symmetric key and the encrypted message) will be delivered to the receiver. 
When received, the receiver first uses her private key to decrypt the symmetric key and then uses this decrypted key to decrypt the encrypted message.  Usually, the email encryption~\cite{poddebniak2018efail} uses such a method to protect the emails from being accessed by unauthorized parties. \looseness=-1 
%Alternatively, the asymmetric access control can also be achieved through access tokens, where the user signs the token, so that the system can verify the token is produced by the authentic user, and restrict the access when the token expires. % Usually, the email encryption~\cite{poddebniak2018efail} uses such a method to protect the emails from being accessed by unauthorized parties.  

\item \textbf{Symmetric Cryptography~\cite{simmons1979symmetric} Based Access Control} is different from asymmetric ones, as it does not use public-private key pairs to protect the resources. Instead, a symmetric key %cryptographic access control protocols usually require 
between the sender and the receiver is exchanged first, and then used for future communications. Since directly using the symmetric keys to communicate may result in the traffic to be recovered through cryptographic analysis, the sender and receiver may use the exchanged symmetric key to derive session keys or tokens. For example, in Bluetooth communication~\cite{zhang2020breaking}, two devices will first negotiate a long term key (LTK), from which to derive the session key every time when establishing a new connection. % that is maintained at both devices, and during each connection, LTK will be used to derive a new session key.    \looseness=-1
\end{packeditemize}
}

\begin{table}
\centering
\scriptsize
 \setlength\tabcolsep{2pt}
\begin{tabular}{lccc}\toprule
                                                            & \textbf{Mini-programs} & \textbf{Mobile apps} & \textbf{Web apps} \\ \midrule
%Install-free?                                               & \tickYes                    & \tickNo                   & \tickYes                  \\ 
\begin{tabular}[c]{@{}l@{}} 
Platforms store users' data on cloud for API use?
\end{tabular}  & \tickYes                    & \tickNo                   & \tickNo                   \\
\begin{tabular}[c]{@{}l@{}} API for social network resources?  \end{tabular}                      & \tickYes                    & \tickNo                   & \tickNo                   \\
\begin{tabular}[c]{@{}l@{}} Nature support of encryption based access protocol? \end{tabular} & \tickYes                    & \tickNo                   & \tickNo                   \\
\begin{tabular}[c]{@{}l@{}}Nature support of token based access protocol? \end{tabular}    & \tickYes                    & \tickNo                   & \tickNo                   \\
\begin{tabular}[c]{@{}l@{}}Nature support of key management protocol?     \end{tabular}        & \tickYes                    & \tickNo                   & \tickNo                   \\
%\begin{tabular}[c]{@{}l@{}}Data or services are consumed~\\on the back-end?   \end{tabular}        & \tickYes                    & \tickNo                   & \tickNo                   \\
\begin{tabular}[c]{@{}l@{}}Apps need to be vetted?  \end{tabular}                                & \tickYes                    & \tickYes                  & \tickNo  \\\bottomrule                 
\end{tabular}
\caption{Comparison of Mini-Programs, Mobile Apps, and Web Apps.  }
\label{tab:iotdevices}
\end{table}

\section{Understanding the Attack Surface}
\label{sec:understanding}

\ignore{
It is obvious that the security of cryptographic access control lies in the management of the cryptographic key (e.g., private key in asymmetric cryptographic access control, or the symmetric key in the symmetric cryptographic access control), which must be carefully protected (e.g., exchanged and maintained securely).
Any mistake in managing the keys can lead to catastrophic consequences. %\looseness=-1
%
%While the mini-programs providers (e.g., \Tencent, \textit{Baidu} and \textit{Alipay}) all have provided similar cryptographic access control protocols,  and briefly discussed how to use the protocols, they all did not explain why the protocol should work in such a way. For example, they only explained the developers should use \token and \appsecret to fetch \sessionkey (details will be given later), and they never explained the purpose of using \token and \appsecret (not \appsecret only). Understanding why is important, as the threat model of mini-programs and mobile apps are different, and understanding the design principles of the protocols will guide the developers to implement their products correctly. For example, 
For instance, a common practice in traditional key management in mobile app development is to hard-code the key in the app's source code~\cite{googlemapkey}, but such practices will be insecure in the mini-programs and many super-app providers explicitly state in their documentation that \appsecret shall not be stored in the mini-program front-end in any forms. 
}

% In this section, we aim to demystify the cryptographic access control in mini-programs, particularly how keys are managed since they are the only secret in modern cryptography, and understand its security weaknesses. %If the developers do not understand the threat model, they can easily make mistakes.
% To this end, we first registered a developer account for each tested super-app,  downloaded the corresponding SDKs and IDEs, and then built mini-programs by following their official documents to set up both its front-end and back-end. To understand how keys are distributed and data are transferred through the traffic, we also reverse engineered the super app code, and used \textsf{Burpsuite}~\cite{wear2018burp} to inspect the network packets. 

In this section, 
%we aim to understand the attack surface in mini-program's sensitive resource access. W
we use \wechat as an example due to its popularity to explain the access control parties (\S\ref{subsec:roles}) and discuss how the access control is implemented (\S\ref{subsec:protocol}). We also provide a threat model of \wechat (\S\ref{subsec:threatmodel}) and possible attacks (\S\ref{subsec:possibleattacks}).
\looseness=-1

% we aim to understand the attack surface in mini-program's sensitive data access. In particular, we use \wechat as an exemplary super-app given its popularity to explain the parties involved in its access control (\S\ref{subsec:roles}) and then discuss in greater detail how the access control is designed and implemented (\S\ref{subsec:protocol}). Finally, we provide the analysis of the possible attacks (\S\ref{subsec:possibleattacks}). %followed by the scope and threat model (\S\ref{sub:scope}). Finally, we describe how to obtain and identify the vulnerable mini-programs (\S\ref{Sub:methodology}).

%and inspected their code.

% Please add the following required packages to your document preamble:
% \usepackage{booktabs}

\subsection{Parties Involved}
\label{subsec:roles}

Sensitive data access is complicated in \wechat mini-programs, since multiple parties including  \wechat client, \wechat server, mini-program front-end, and mini-program back-end all need to be involved, as shown in \autoref{fig:arch}. \looseness=-1

%In a \wechat mini-program, there are four parties involved when accessing sensitive data, as shown in \autoref{fig:arch}, in the mini-program paradigm of \wechat, four parities can be involved. (1) \wechat client provides the running environment for all the mini-programs; (2) \wechat server offers the client and mini-programs various services (\eg, vetting services, sensitive data fetching); (3) mini-program frontends   run atop the \wechat client; (4)  back-ends   communicate with the frontends, and provide various services such as storage.  \looseness=-1

\begin{itemize} 
\item \textbf{\wechat Client (WC)}  provides a JavaScript runtime environment that enables developers to write and run their mini-programs using JavaScript. Moreover, it also offers a range of APIs that support a wide range of needs in mini-program development, including UI rendering, resource access, and network accessing.

%extend their functionalities of the mini-programs. These APIs can be used to render the UIs, fetch system resources and communicate with its back-end.  
\item \textbf{\wechat Server (WS)} is crucial in the \wechat mini-program ecosystem. They vet submitted mini-programs (similarly to how Google Play vets Android apps), and securely store and deliver sensitive resources (e.g., services and data) to trusted mini-programs upon request. \looseness=-1

%plays a vital role in the \wechat mini-program ecosystem. In addition to vetting each mini-program submitted by the developers (similarly to how Google Play vets Android apps), WS also needs to store the collected sensitive resources (\eg, service) and deliver them securely to trusted mini-programs when requested. \looseness=-1

%Basically, \wechat server needs to tackle the mini-program reviewing process (\ie, all the mini-programs are reviewed by the \wechat server before being published), and more importantly, as discussed, it has to deal with the sensitive data transmission, since it maintains the \wechat specific user sensitive information.  \looseness=-1

%For example, \wechat does not allow the user's \wechat password to be collected by the mini-programs, and mini-programs that have such behaviour will be rejected by \wechat server. 
%Second, as discussed, \wechat can provide various sensitive information that collected by \wechat. 
\item \textbf{Mini-program's Front-end (MF)} handles UI rendering, user requests processing, and communicates with the mini-program back-end to execute specific business logic like online shopping and ride-hailing. Unfortunately, MF is developed using JavaScript, making it susceptible to reverse engineering.

%As a core component of mini-program, the mini-program front-end runs atop of the JavaScript environment provided by \wechat. Since the size of the mini-program front-end is usually a few MB (\wechat requires the mini-program packages cannot cannot exceed 12 MB~\cite{sizelimit2021}), the code complexities of these mini-programs are relatively low. As such, it would be easy for the hackers to reverse engineer these mini-programs and perform attacks against these mini-programs. \looseness=-1 %\wechat has adopted a series of mitigation to prevent the code from being reverse engineered. For example, obfuscation is enabled by default, and for the mini-programs running on Windows, \wechat even enforced the package encryption.

%These frontends are compiled, compressed, and distributed as \texttt{.wxapkg} file. Similar to the Android native apps, each \texttt{.wxapkg} file contains a configuration file named \texttt{app.json}, the mini-program version of \texttt{Mainfest}, containing the authorization scope information and other configuration information. Moreover, each \texttt{.wxapkg} file also contains one or more the JavaScript code files, which are the mini-program version of the source code files, and multiple page files, which are the mini-program version of UI layout files. 
\item \textbf{Mini-Program's Back-end (MB)} 
 is used to store user specific data related to the mini-program. 
 These MBs are the backbones of mini-programs that handle both user data generated from front-end, and user sensitive resources delivered by \wechat severs. % delivered sensitive data process them securely with cryptographic a. % Recall that the front-end of mini-programs can be hacked easily, and therefore, \wechat requires that the security related functionalities
 %such as key management and data decryption. % should be preformed at the back-ends. 

\end{itemize}

\begin{table}[t]
\centering
 \belowrulesep=0pt
 \aboverulesep=0pt
 \setlength\tabcolsep{3pt}
\scriptsize{
\begin{tabular}{@{}llccll@{}}
\toprule
  \textbf{Definition} & \textbf{\begin{tabular}[c]{@{}c@{}}Length\\ (bits)\end{tabular}} & \textbf{\begin{tabular}[c]{@{}c@{}}Generated\\ by\end{tabular}} & \multicolumn{1}{c}{\textbf{\begin{tabular}[c]{@{}c@{}}Involved\\ Parties\end{tabular}}} & \multicolumn{1}{c}{\textbf{\begin{tabular}[c]{@{}c@{}}Validation\\ Period\end{tabular}}} \\ \midrule
      Master Key (\appsecret)         & 256                                                              & WS                                                              & MB, WS                                                                                &         $\infty$                                                                                    \\
  Encryption Key (\sessionkey)      & 128                                                              & WS                                                              & MB, WS                                                                                &     User-specific                                                                                     \\
Login Token (\token)         & 128                                                              & WS                                                              & WS, MB, WC, MF                                                                        &     5 mins                                                                                     \\ 
         Access Token (\texttt{AT})        & 512                                                             & WS                                                              & WS, MB                                                                  &     120 mins                                                                                     \\ 
\bottomrule
\end{tabular}
 
%%%%%%%%%%%%%%%%%%%%%%%%%%%%%%%%%%%%%%%%%%%%%%%%%%
}
\smallskip
\caption{Summary of various keys used in \wechat ecosystem.}
\label{tab:symbols}
\end{table}

\subsection{Cryptographic Access Control}
\label{subsec:protocol}

The \appsecret is a vital key for cryptographic access control in mini-programs, generated by \wechat upon authentication. \wechat offers two \appsecret-based protocols for accessing resources: encryption-based for sensitive data and token-based for sensitive services.

% The master key (\appsecret) is the foundation of the cryptographic access control in mini-programs. In particular, once a mini-program is authenticated, \wechat server generates its \appsecret (a 256-bit cryptography key, and \wechat document also calls it \texttt{AppSecret}), and distributes it to the mini-program back-end. 
% Note that each mini-program can only have one unique \appsecret. As such, \wechat uses the \appsecret to uniquely authenticate each specific mini-program. As shown in \autoref{tab:symbols}, since \appsecret never expires,  once \appsecret gets lost, the mini-program developer has to ask \wechat again and obtain a new \appsecret, and all the legacy sessions that involved the old \appsecret will be considered invalid.  
% As there are two types of resources that can be accessed by the \appsecret, there are two types of protocols that are available accordingly: the encryption-based access control for sensitive data and the token-based access control for sensitive services. 

  \begin{figure}[t]
    \centering
    
     \includegraphics[width=\linewidth]{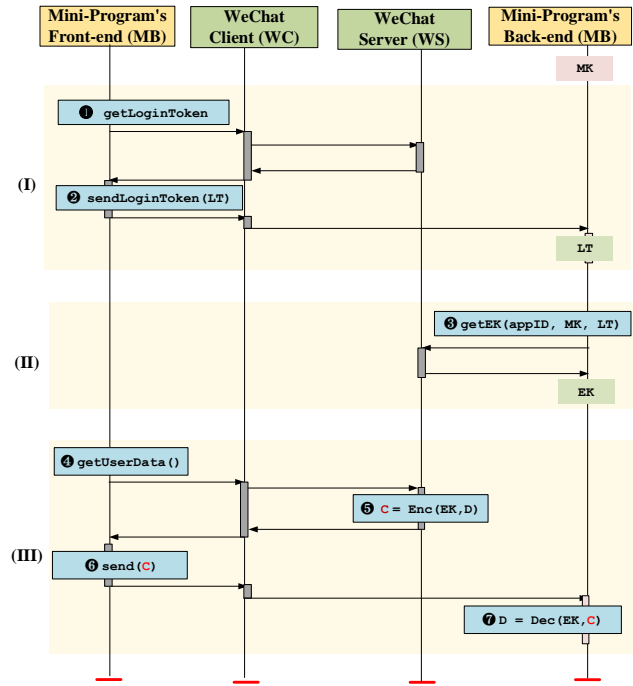}
     \vspace{-0.3in}
    \caption{Sensitive data access protocols in \wechat mini-programs}
    \label{fig:arch}
     
\end{figure}

\paragraph{Encryption-based Access Control} 
To safeguard sensitive user data from abuse or manipulation, \wechat encrypts it when accessed by mini-programs through APIs.  The encrypted data is  sent back to the mini-program's back-end for decryption and processing, following a three-phase process shown in \autoref{fig:arch}:

\ignore{
Note that there is no public documentation describing this protocol, and we obtain it through our manual reverse engineering and testing. \looseness=-1

\begin{itemize} 
\item \textbf{\wechat Client (WC).} \wechat client is the host mobile app that provides the JavaScript runtime environment as well as the programming APIs for the mini-programs. Note that \wechat mini-programs are programmed using JavaScript, and these JavaScript APIs support most of the needs in mini-program development such as UI rendering, data access, and network communications. 

%extend their functionalities of the mini-programs. These APIs can be used to render the UIs, fetch system resources and communicate with its back-end.  
\item \textbf{\wechat Server (WS).} \wechat server plays a vital role in the \wechat mini-program ecosystem. In addition to vetting each mini-program submitted by mini-program developers (similarly to how Google Play vets Android apps), \wechat server also needs to store the collected sensitive user data and deliver them securely to trusted mini-programs when requested.

%Basically, \wechat server needs to tackle the mini-program reviewing process (\ie, all the mini-programs are reviewed by the \wechat server before being published), and more importantly, as discussed, it has to deal with the sensitive data transmission, since it maintains the \wechat specific user sensitive information.  \looseness=-1

%For example, \wechat does not allow the user's \wechat password to be collected by the mini-programs, and mini-programs that have such behaviour will be rejected by \wechat server. 
%Second, as discussed, \wechat can provide various sensitive information that collected by \wechat. 
\item \textbf{mini-program Front-end (MF).} The front-end of a mini-program is responsible for rendering the UI, processing user request, and communicating with the corresponding mini-program back-end for specific business logics such as online shopping, appointment booking, and ride hailing. Unlike Android apps that can have almost unlimited sizes, the front-end of the mini-program is limited to maximum 12 MB~\cite{sizelimit2021}. Since they are developed with JavaScript, they can be easily reverse engineered. \looseness=-1

%As a core component of mini-program, the mini-program front-end runs atop of the JavaScript environment provided by \wechat. Since the size of the mini-program front-end is usually a few MB (\wechat requires the mini-program packages cannot cannot exceed 12 MB~\cite{sizelimit2021}), the code complexities of these mini-programs are relatively low. As such, it would be easy for the hackers to reverse engineer these mini-programs and perform attacks against these mini-programs. \looseness=-1 %\wechat has adopted a series of mitigation to prevent the code from being reverse engineered. For example, obfuscation is enabled by default, and for the mini-programs running on Windows, \wechat even enforced the package encryption.

%These frontends are compiled, compressed, and distributed as \texttt{.wxapkg} file. Similar to the Android native apps, each \texttt{.wxapkg} file contains a configuration file named \texttt{app.json}, the mini-program version of \texttt{Mainfest}, containing the authorization scope information and other configuration information. Moreover, each \texttt{.wxapkg} file also contains one or more the JavaScript code files, which are the mini-program version of the source code files, and multiple page files, which are the mini-program version of UI layout files. 
\item \textbf{mini-program Back-end (MB).} 
 \wechat mini-programs can also have a back-end to store user specific data related to the mini-program. 
 These mini-program back-ends are the backbones of mini-programs that handle both user data generated from the front-end, and also user sensitive data delivered by \wechat severs. % delivered sensitive data process them securely with cryptographic a. % Recall that the front-end of mini-programs can be hacked easily, and therefore, \wechat requires that the security related functionalities
 %such as key management and data decryption. % should be preformed at the back-ends. 

\end{itemize}

}

\begin{enumerate}[label=(\Roman*)]

\item \textbf{Login Token (\token) Acquisition.} 
\wechat assigns a unique appID and \appsecret to each mini-program to verify their identity and developers. However, since sensitive data is linked to users rather than mini-programs, appID and \appsecret alone are insufficient. To ensure users only access their data, \wechat generates a user-specific \token when they log in, which is combined with the appID and \appsecret to obtain the \sessionkey (\textbf{Step \ding{182}}). As shown in \autoref{tab:symbols}, \token is a 128-bit hex string valid for only five minutes~\cite{wechatkey20}, generated during user login to the \wechat server. MF delivers it to MB for querying \sessionkey within the time window (\textbf{Step \ding{183}}). This design prevents unauthorized collection of user data at scale by MB.

\item \textbf{Encryption Key (\sessionkey) Fetching (\ding{184}).} \sessionkey encrypts user data and requires appID, \appsecret, and \token for retrieval. To balance protection against enumeration attacks and mini-program data processing performance, the \sessionkey has a dynamic expiration period, set to five minutes by default~\cite{wechatkey20}. Its expiration time may be extended or reduced based on API usage frequency. Although \token is discarded after obtaining the \sessionkey, there's a possibility of two separate \sessionkey acquisitions returning the same \sessionkey within the expiration period, despite different \token{s}.\looseness=-1

% \sessionkey is a user-specific cryptographic key that is used to encrypt the sensitive data collected at \wechat servers before sending to the mini-program server, and the mini-program server requires this key to decrypt the sensitive data. To request the user specific \sessionkey, the mini-program back-end needs to provide three parameters: the \token obtained in stage-II, \appsecret obtained in stage-I, and the mini-program ID (\textbf{Step \ding{185}}). %If all the three inputs are provided correctly,  
% Then the \wechat server will send the \sessionkey, if it has not expired yet, to the mini-program back-end; otherwise the server will generate a new \sessionkey and send it to the back-end.
% According to the documentation, the period of validity of an \sessionkey is not fixed. Instead, it changes dynamically depending on how often the mini-program uses them. The more popular a mini-program is, the longer the period of validity of its \sessionkey will be~\cite{sessionkey}. Also, similar to \appsecret, \sessionkey should also not be distributed to the front-end in any cases. 

\ignore{
which is generated based on the three inputs. Note that \sessionkey can be generated before the mini-program's back-end sends the request. If this is the case, \wechat server will return the existing \sessionkey if \sessionkey has not expired.  
 As shown in \autoref{tab:symbols}, another interesting fact is that the period of validity of a \sessionkey is not fixed. Instead, it changes dynamically depending on how frequent the mini-program get used. If the mini-program is popular, the period of validity of its \sessionkey will be longer~\cite{sessionkey}. 
 The back-end then saves \sessionkey onto its disks.
 Recall that in the \wechat threat model, the front-end cannot be trusted, and therefore, similar to \appsecret, \sessionkey should also not be distributed to the front-end in any cases. 
 }

  \begin{figure}[t]
     \centering %\vspace{-0.1in}    
      \includegraphics[width=0.92\linewidth]{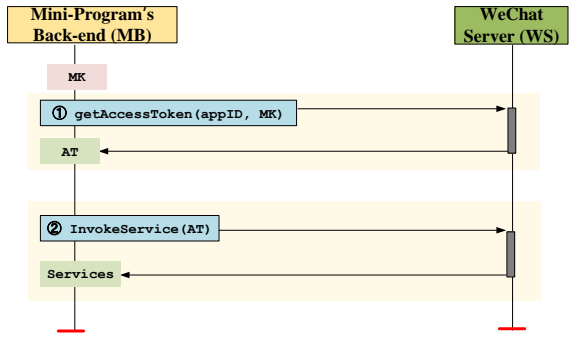}
 
    \caption{Cloud services access protocols in \wechat mini-programs}
    \label{fig:services}
    
\end{figure}

\item \textbf{Sensitive Information Encryption and Decryption.} 
 MF uses data fetching APIs to retrieve user data (\eg, \texttt{getPhoneNu\\mber}) from WS (\textbf{Step} \ding{185}). WS identifies the user and mini-program from WC's request, encrypts sensitive data using the corresponding \sessionkey from stage-II, and sends encrypted data to WC (\textbf{Step} \ding{186}). WC forwards the encrypted data to its back-end (\textbf{Step} \ding{187}). Decryption should only occur at the back-end using the \sessionkey obtained in stage-III, since the front-end is untrusted. The back-end decrypts the data (\textbf{Step} \ding{188}) to obtain the plaintext of sensitive data. The MB may then use this data for specific business logic, such as retrieving app-specific data based on decrypted phone number. %we find that many of the back-ends use the users' phone number to retrieve its users. \looseness=-1

 %\end{packeditemize}
\end{enumerate}

% \paragraph{Sensitive Service Access (via Access Token \textsf{AT})}  
%  \begin{figure}
%     \centering
    
%      \includegraphics[width=0.95\linewidth]{image/twocases.pdf}
%      \vspace{-0.1in}
%     \caption{Service consumption under benign and malicious cases}
%     \label{fig:twocases}
    
% \end{figure}

\paragraph{Token-based Access Control} 
 \wechat platform has provided various sensitive services that encapsulate complex functionalities such as AI chat bot or OCR for mini-program developers to purchase and use. As shown in \autoref{fig:services}, the use of these services is quite straightforward, % given that these services are mini-program-specific (\wechat will charge the mini-programs that used those services). 
 %as shown in \autoref{fig:services}, and it
 which is a two-step process: (i) the MB provides the appID and the mini-program \appsecret to the WS in exchange for API access token \textsf{AT} (Step \ding{192}); (ii) then, when invoking services provided by \wechat, the MB  has to attach both \textsf{AT} and the data sent to WS (Step \ding{193}). As shown in \autoref{tab:symbols}, \textsf{AT} is 512 bits long and valid for 2 hours. \looseness=-1

\ignore{
\subsection{{\bf Case Studies}}
\label{sub:attack}
%In this section, we will perform case studies on the developers of vulnerable mini-programs, reactions to the vulnerabilities, as well as an analysis of the impacts that may be caused by the leaked \appsecret.

 {According to how the mini-programs utilize the sensitive data, different mini-programs may expose themselves to different attacks when their \appsecret{s} are leaked. In general, we conduct two types of case studies corresponding to our two types of attacks to show how the attacker can manipulate the predictable data (one case study) and the non-predictable data (three case studies) to cause the serious impacts against their users. Particularly, we would like to present the implementation details in the first case study with other case studies briefly discussed, since the implementation processes of the later three case studies are similar to the first one.   }

\subsubsection{Manipulating  Predictable  Data (i.e., users’ phone numbers)} Recall that the  predictable  data  can be used as the primary key to retrive the user information, and if leaked, the attacker can mount the account hijacking attacks. Next, we present a concrete account hijacking attack  % via phone number manipulation 
against \texttt{Nestle (We Proudly Serve)} mini-program to further demonstrate the consequence of the attack. %real impacts. %Before our attack, we first unpack the front-end of \texttt{StarBucks} mini-program using \texttt{WXUnpacker}~\cite{decomplier}, and extract \appsecret. We save the \appsecret for further usage. 
While performing our attack, we first registered two accounts: a victim account and an attack account, with \wechat, and use the mini-program with these two accounts, as we must make sure we attack our own account, a community practice for the attack work. % did take the ethics into the highest standard, and therefore, we only launched the attacks against our own accounts (both the victim account and the attack account discussed below are our own account). 
As described in \S\ref{sub:overview:attack}, the attack involves two stages: (I) Obtaining attacker’s encryption key (\sessionkey), and (II) User phone number retrieval and manipulation.

\paragraph{(I) Obtaining Attacker’s Encryption Key (\sessionkey)} In this stage, we first get \token from the \wechat server and use the obtained \token, Mini-program ID, and the leaked \appsecret to query the encryption key \sessionkey.  To get the \token from the \wechat server, we use a well-known MITM proxy \texttt{Burp Suite}~\cite{wear2018burp} to inspect the traffic between the \wechat client and its server.  By default, when a victim mini-program invokes \texttt{wx.login},  the \token will be delivered to the front-end and require the front-end to send it to the back-end. In our attack, we intercepted the \token (i.e., the \texttt{encrypt\_code} in \autoref{fig:attackexample})  without sending it to the back-end as shown in  \autoref{fig:attackexample} (\textbf{Step \ding{182}}).

Next, we programmed a python script to query the \sessionkey from \wechat server, and the obtained \token is then fed to the \texttt{getEK()} API.  As shown in \autoref{tab:inner:apis1}, it can be observed that \texttt{getEK()} takes three inputs, including the public accessible appID,  the \texttt{secret} (\ie, \appsecret), and  \texttt{Js\_code} (\ie, the \token obtained in \textbf{Step \ding{182}}). As shown in \autoref{fig:attackexample}, \wechat server returns us the encryption key \sessionkey ``\breakabletexttt{a9ah7ZiIDSxZU6oTzLCW6g==}'', which is encoded using Base64 (\textbf{Step \ding{183}}). \looseness=-1

\begin{figure*}[th]
    \centering
    
    \includegraphics[width=0.95\linewidth]{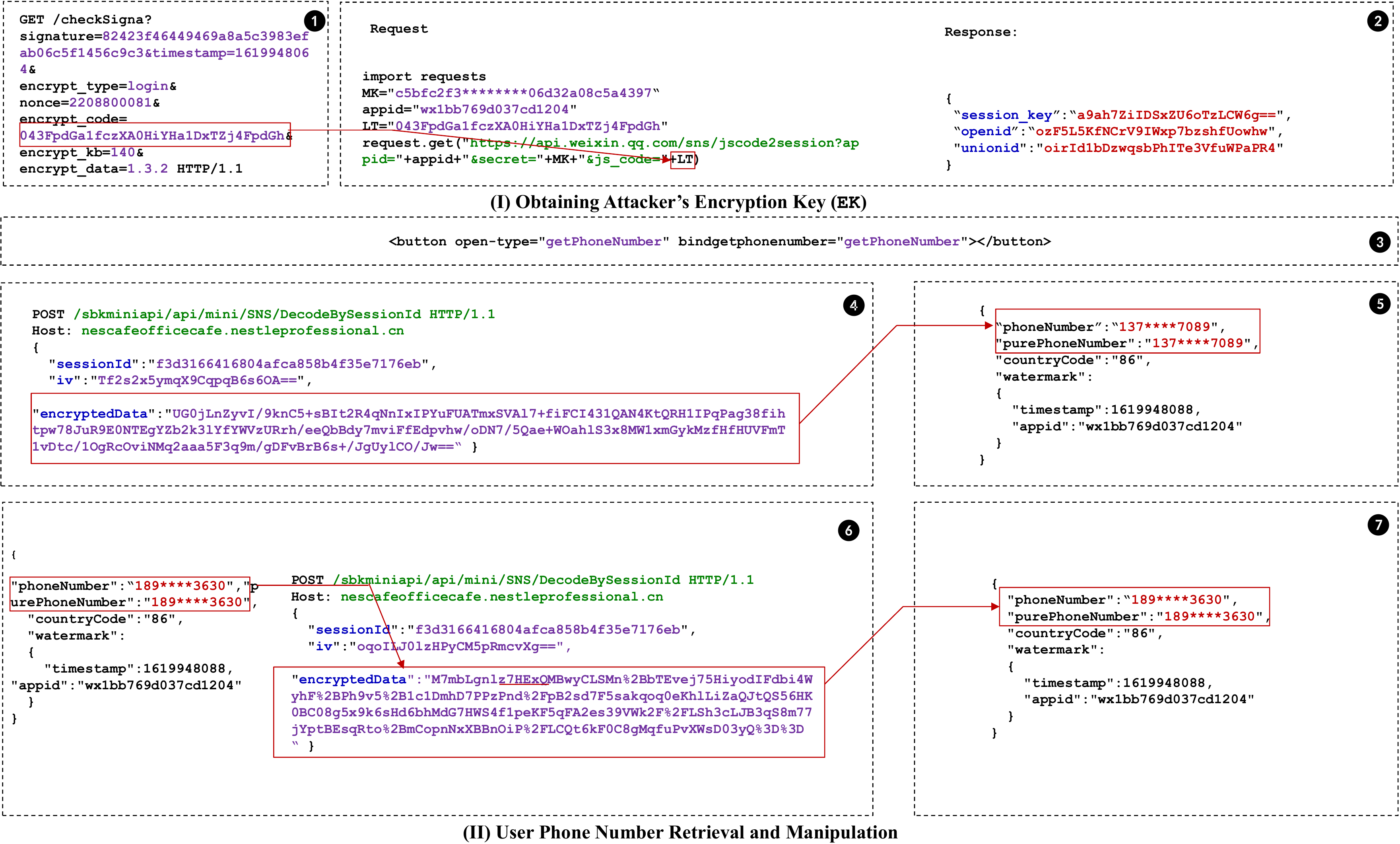}
    \caption{An excerpt of attack traffic traces}
    \label{fig:attackexample}
       
\end{figure*}

% Recall that the obtained \sessionkey can get expired, and the validity time of \sessionkey depends on how often the user uses the mini-program. Therefore, to check the validity time of \sessionkey, we performed an analysis on the validity time of \sessionkey. Specifically, we invoke \texttt{getEK()} multiple times with different intervals (\ie, 2$s$, 10$s$, 60$s$, 120$s$), and check at what frequency, \wechat server will change their \sessionkey. We present the results in \autoref{fig:stat_2s},  \autoref{fig:stat_10s},  \autoref{fig:stat_60s} and \autoref{fig:stat_120s}. Surprisingly, we found that \sessionkey remains the same (\ie, around 2 minutes) regardless of how often we invoke \texttt{getEK()}.   This is inconsistent from \wechat official document. By knowing it, we can query the \sessionkey every 2 minutes to ensure the freshness of our \sessionkey. 

%at a fixed interval between 2 seconds to 120 seconds, and the results are shown in \cref{fig:stat_2s} to \cref{fig:stat_120s}. As plotted in the figures, the \sessionkey of a mini-program remains to expire within 120 seconds when mini-program probes the \wechat server from 2 seconds a time to 2 minutes a time. Therefore, an attacker can immediately trigger another log in to renew the \appsecret within 2 minutes to perform enumeration attack with intercepted and decrypted network requests. 

\paragraph{(II) Predictable Data Retrieval and Manipulation} In this stage,  we intentionally trigger  \texttt{We Proudly Serve}'s front-end to invoke \texttt{getPhoneNumber}, and manipulate the encrypted phone number sent from the \wechat server to trick the back-end of  \texttt{We Proudly Serve} into believing it is interacting with the victim's account.
%
%This is because, like many other vendors, \texttt{StarBucks} leverages the user's phone number to retrieve and uniquely associate with their users. 
%Recall that we have two accounts, one is assumed to belong to a victim, and another is assumed to belong to the attacker. Therefore, if we can modify the attacker's phone number to the victim's, \texttt{StarBucks} will believe that it interacts with the victim but not the attacker. Specifically, we used five steps to modify the phone number. 
More specifically, we first trigger  \texttt{We Proudly Serve} to invoke \texttt{getPhoneNumber}. As shown in \textbf{Step} \ding{184} of \autoref{fig:attackexample}, \texttt{getPhoneNumber} is the callback function of the button, and when we press it, it gets called. Then we intercept the cipher that is about to be sent to the mini-program back-end server (\textbf{Step \ding{185}}), and this cipher was initially received by the mini-program from \wechat server when invoking the \texttt{getPhoneNumber} API. 

%sent from the front-end using the same method above (\textbf{Step \ding{185}}). 

According to \wechat's official document~\cite{wechatkey20}, \wechat uses the standard cryptographic algorithms (\ie,  the data encryption and decryption algorithm is AES-128-CBC~\cite{bulens2008implementation} and the data padding algorithm is PKCS\#7 \cite{kaliski1998pkcs}), and therefore, we can easily implement our own script to decrypt and encrypt the {\tt encryptedData} % encryption and decryption algorithms using python3.  Therefore, at this moment, we decrypt the cipher 
using the obtained \sessionkey,  and inspects its format, so that we can forge fake phone number in the following steps (\textbf{Step \ding{186}}). After inspecting the format of the decrypted plain-text, we can then replace the attacker's phone number  (\ie,\breakabletexttt{137****7089}) with our victim's phone number (\ie, \breakabletexttt{189****3630"}), and encrypt the modified phone number using \sessionkey  (\textbf{Step} \ding{187}). 
As discussed, the \sessionkey used by the backend of \texttt{We Proudly Serve} to decrypt the phone number and \sessionkey used by the attacker to encrypt the phone number are the same,  and therefore, our modified phone number will be successfully decrypted by the back-end of \texttt{We Proudly Serve} (\textbf{Step} \ding{188}), and used by \texttt{We Proudly Serve} to retrieve the user information.  It is interesting to note that much of the retrieved information is collected and maintained by \texttt{We Proudly Serve}, meaning that if we hijack multiple accounts by enumerating all the phone numbers, we can harvest a large amount of \texttt{We Proudly Serve}'s user information.  \looseness=-1

\subsubsection{Manipulating  Non-Predictable  Data} We have already demonstrated how an attacker can mount account hijack attacks via the predictable data (i.e., phone number), and in the following, we would like to illustrate how the attacker can manipulate the non-predictable data. Particularly, we would like to demonstrate how an attacker can manipulate the WeRun and User Information.
% Recall that there are seven types of sensitive data that can be collected by the mini-program. Given that we have already demonstrated how to explore the phone number, and we have not found any mini-programs that use group chat info and address info, in the following we present two case studies corresponding to the these two types of sensitive data.
% 7-1-2=2, Missing shareInfo and privateMessage
%\ZQ{}

%Recall that there are seven types of sensitive data that can be collected by the mini-program. Given that we have already demonstrated how to explore the phone number, and we have not found any mini-programs that use group chat info and address info, we will present two case studies corresponding the left two types of sensitive data. 

\begin{itemize}
    % \item \AY{User Phone Access. Since user phone number are often used as identifiers to associate mini-program users with that of native application, the phone number should be carefully protected. However, we found that there are many mini-programs that use user phone number to have leaked the \appsecret, which enables attackers to change the phone number used for logging in arbitrarily to fake their identities. In the next section, we will provide a concrete example on how this case poses threat both on mini-program users and developers.} \ZY{This case should be removed, as we have already discussed the possibilities.}
    
    \item  \textbf{\texttt{WeRun} Data Manipulation for Cheating and Profit.} {\texttt{WeRun} data contains the steps a user walks each day. Among the vulnerable mini-programs, we observed that some mini-programs use this feature to monitor the sports condition of users, such as universities (e.g., Beihang University)  that require minimal daily sports for students, and these daily sports will be used to grade the students. As a student,  he or she can follow the steps \ding{182}--\ding{188} that introduced in our previous case study to manipulate the daily step counts and achieve cheating.  
    Some mini-programs (e.g., Zhongchuang Steps \& Entertainment) %advocates users to protect the environment by allowing users to 
    reward their users based on the user's daily step counts. As such, a malicious user may also exploit the mini-programs by manipulating the steps to earn the rewards.} \looseness=-1
    \item \textbf{User Information Access for Account Information Stealing}  The \texttt{getUserProfile} API enables mini-program developers to acquire the basic account information such as avatars, gender, and nicknames. However, the information can also be collected by attacker if the \appsecret is leaked, since the attacker can use the steps \ding{182}--\ding{186} to decrypt the encrypted user information, and send the decrypted data to a remote server. For example, we confirmed that \texttt{Cool Run Everyday} from \Tencent themselves leaked their \appsecret, enabling attackers to collect sensitive the user information.
    
    %attackers can fake their nicknames and avatars, especially in social mini-programs or games. For instance, we confirmed that \texttt{Cool Run Everyday} from \Tencent themselves leaked their \appsecret, enabling attackers to change the nickname arbitrarily. As such, the attackers can utilize social engineering techniques to fake close friends of playeers to make money. \ZY{I think the motivation here is very weak and does not make senses at all. Why the attacker use such method to attack}
    % \item \AY{Message Forwarding.}
\end{itemize}

\ignore{\subsection{Developer Responses of \appsecret leakage}
\AY{Prior to our large-scale measurement analysis on the 2 million mini-programs, we manually identified \largecompanycases mini-programs from large vendors. However, different from native application stores such as Google Play, the mini-program metadata does not contain developer information
and we immediately Before we are able to collect all the 2 million mini-programs, we performed a perliminary analysis by sampling mini-programs, and we identified that 37 mini-programs considered to be related to large companies in Fortune top 500 list, as shown in \cref{tab:appsecret}. After we identified these vulnerable mini-programs and confirmed the vulnerability, we reported our findings to \Tencent, and received their confirmation. As \Tencent agreed on notifying the developers of their \appsecret leakage, we would like to investigate how the developers responded to this exposure. As such, we re-collected the packages of these mini-programs 2 months after our initial disclosure to see if the original leaked \appsecret{s} are removed, refreshed, or remained as they were. As shown in \cref{tab:appsecret}, we found that 11 of these mini-programs removed their leaked \appsecret{s} to fix the problem, where as 18 of them, though intended to fix the problem, took the wrong way: they merely refreshed their \appsecret{s}, which still exposes their mini-programs in danger. We are surprised to find that there are still 8 mini-programs did not take any actions at all but kept their original exposed \appsecret{s} in use, which definitely makes their mini-programs exploitable by malicious attackers, as soon as the attackers obtains the \appsecret. In the next sections, we will provide a detailed explanation of how such an exposure will lead to severe consequences by malicious mini-program users.}}
}

\begin{table}[]
\scriptsize
\setlength\tabcolsep{2pt}
\begin{tabular}{@{}llll@{}}
\toprule
\textbf{Keys} & \multicolumn{1}{c}{\textbf{Goals}}                                          & \textbf{Protected Assest} & \textbf{Default Assumptions}                                                         \\ \midrule
MK            & Querying EK/AT                                                              & EK,and AT                 &  Developers keep \appsecret{s} within MSs                                                              \\
EK            & Encryption and Decryption                                                   & Sensitive Resources       & \begin{tabular}[c]{@{}c@{}}Developers fetch \sessionkey{s} within MSs \end{tabular} \\
LT            & \begin{tabular}[c]{@{}l@{}}Authenticating User; \\ Querying EK\end{tabular} & EK                        & \begin{tabular}[c]{@{}c@{}} Developers fetch \token within MFs\end{tabular}      \\
AT            & Consuming Services                                                          & Paid Services             & \begin{tabular}[c]{@{}c@{}}Developers fetch \textsf{AT}{s} within MSs\end{tabular} \\ \bottomrule
\end{tabular}
\caption{Threat model of \wechat}
\label{tab:threatmodel}
\end{table}

% Please add the following required packages to your document preamble:
% \usepackage{booktabs}

\begin{table}[]
\scriptsize
 
 \setlength\tabcolsep{2pt}
\begin{tabular}{@{}lccc@{}}
\toprule
\textbf{Keys} & \textbf{Data}             & \textbf{Services}           & \textbf{Feasible} \\ \midrule
\appsecret            & Manipulating Data (\tickcoloredYes)         & Consuming Services for Free (\tickcoloredYes) & \tickYes          \\
\sessionkey           & Manipulating Data  (\tickcoloredYes)        & N/A                         & \tickYes          \\
\token            & Collecting Other's Data (\tickcoloredNo) & N/A                         & \tickNo           \\
\texttt{AT}            & N/A                       & Consuming Services for Free (\tickcoloredYes) & \tickYes           \\ \bottomrule
\end{tabular}
\smallskip
\caption{Summary of possible attacks. ``\tickcoloredYes'' means the corresponding attack works, while ``\tickcoloredNo'' means does not.  }
\label{tab:attackspossible}
\end{table}

\subsection{Threat Model of \wechat}
\label{subsec:threatmodel}

\wechat has defined four types of keys, each with a distinct purpose, and \wechat has made different assumptions about the security of these keys. To clarify this, we summarize the threat model to highlight \wechat's use and default assumptions for each key, as shown in \autoref{tab:threatmodel}. 

\begin{itemize}
    \item \textbf{\appsecret
} serves as the root of WeChat security and is used for querying EK and AT. \wechat assumes that the MK should be strictly held within the MS and never disclosed to MF or any other parties. 

\item \textbf{\sessionkey}
 is responsible for encrypting and decrypting sensitive information, such as phone numbers. It is queried by MS and kept there until it expires.  \wechat assumes that only MS, and not MF, can retrieve the \sessionkey and that all encrypted data must be consumed at MS using the \sessionkey.
 %Since MS is out of reach of potential attackers, \wechat assumes that attackers cannot hack into MS to obtain EKs for the purpose of data manipulation.  
 
\item \textbf{\token
} is crucial for authenticating a user, as it is user-specific and only the data owner can produce a valid \token based on the user's credentials. Additionally, \token is used as the index for querying the session key. \wechat assumes that attackers cannot obtain LT, as doing so would require them not only to install malware, but also to obtain root privileges to access \wechat's data.

\item \textbf{\textsf{AT}} is a critical key for enabling the consumption of paid online services. It is queried by MS and remains stored there until it expires, typically after two hours. \wechat assumes that the consumption of services using AT is carried out by MS instead of MF. \looseness=-1

\end{itemize} 

\subsection{Attack Surface Analysis}
\label{subsec:possibleattacks}

While developers are responsible for managing the \appsecret, some may not adhere to \wechat's development guidelines~\cite{appsecret}. This can result in developers hardcoding or distributing keys to the wrong party, such as distributing MS keys to MF, leading to key leaks to attackers. Violation of the underlying assumptions can result in possible attacks, as shown in \autoref{tab:attackspossible}. 

\begin{itemize}

\item \textbf{With the Leak of \appsecret.} Developers may hardcode the \appsecret in MF, creating opportunities for attackers to obtain it through unpacking the MF.
 If the \appsecret is leaked, attackers can always obtain the \sessionkey to encrypt and decrypt data, compromising confidentiality and integrity. They can also use the \appsecret to obtain an access token and consume services for free. As such, the first assumption made by \wechat may not always hold true. \looseness=-1

% Since it is completely up to the developers to manage \appsecret, not all of them will follow \wechat's development guidelines.
% Similarly to traditional API keys that are usually hardcoded in the front end~\cite{googlemapkey}, mini-program developers can hardcode \appsecret in MF. With the leaked \appsecret, attackers can 
% always obtain \sessionkey to encrypt and decrypt the data, breaking both confidentiality and integrity (\eg, causing data leakage and data manipulation attacks). Since \texttt{AT} can be fetched by \appsecret as well, the attacker can also use \appsecret to query a valid \texttt{AT} to consume the services for free.   
  
\item \textbf{With the Leak of \sessionkey.} It could be possible that careless mini-program developers can accidentally send the \sessionkey obtained on MB to MF. If attackers obtain the \sessionkey, again, the attacker can launch data manipulation attacks, although this attack can last for a short period of time in a particular session and the attacker has to keep refreshing \sessionkey from the mini-program servers. In addition, in the event that the \appsecret is leaked, the attacker is able to obtain \sessionkey through the encryption-based access control protocol, though this would be limited to the attacker's own \sessionkey. As a result, the second assumption made by \wechat may not always hold true. \looseness=-1
 
\item \textbf{With the Leak of \token.} Since \token is linked to an individual user, attackers can create a malicious mini-program with their own \appsecret to retrieve and decrypt all sensitive information of a user who leaks their \token. However, obtaining the \token is difficult, as user authorization is required explicitly. Though attackers can try to hack into the user's account or break \wechat's mechanism to generate a valid \token associated with a specific user, these methods are not practical. Therefore, while attackers can obtain their own \token, they cannot obtain others' \token. Third assumption made by \wechat remains valid.

\item \textbf{With the Leak of \texttt{AT}.} 
If attackers obtain an \texttt{AT}, they can misuse it for up to two hours before it expires. Within this timeframe, they can exploit the compromised \texttt{AT} to consume services that may not be free, resulting in financial charges against the developers. Although it may not be practical for attackers to obtain an \texttt{AT} by compromising the back-end of the mini-program, developers may mistakenly  distribute \texttt{AT}s to MF, and attackers can also query \texttt{AT}s using a leaked \appsecret. As a result, the fourth assumption made by \wechat can also be compromised.
\end{itemize}

\section{Measuring the \appsecret  Leaks}
\label{sec:measurement}

Our investigation confirmed that the keys could be leaked in the front-end of a mini-program, as revealed by the attack surface (\S\ref{subsec:possibleattacks}). However, it remains unclear how widespread key leaks are among mini-programs. Therefore, we have conducted a measurement study to answer the following research questions:

\begin{itemize}
\item \textbf{RQ1:} What are the categories of the \appsecret leaked mini-programs?
\item \textbf{RQ2:} What are their ratings?
\item \textbf{RQ3:} What are the their accessed resources?
\item \textbf{RQ4:} Who are their developers?
\item \textbf{RQ5:} When are their latest update?
\item \textbf{RQ6:} Are there any high profile \appsecret leaked mini-programs?

\end{itemize}

\ignore{
\subsection{Scope and Threat Model} 
\label{sub:scope}

\paragraph{Scope} In this study, we focus specifically on understanding the protection provided by \appsecret-based access protocols in the mini-program ecosystem. While there are other types of attacks that can be carried out using leaked keys, we focus on systematically studying \appsecret leaks to understand their prevalence, root causes, and consequences. We do not aim to detect attacks that are caused by leakage of the \sessionkey, \token, or \textsf{AT}. We use the example of \wechat to explain \appsecret-based access control protocols, but note that similar protocols are used on other mini-program platforms, such as \textsf{Baidu}~\cite{baidusmartprogram} and \textsf{Alipay}. We focus on \wechat in particular due to its popularity and support for both sensitive data access and cloud service access.

%, and it supports the mini-programs to use \appsecret to fetch both \sessionkey and \texttt{AT} for resources consumption, while others such as \textsf{Baidu} only support the sensitive data consumption.  

%  In the communication protocol between mini-programs and social network app, there are mainly three parties: the vendor, the user, and the social network app. The social network app executes mini-program front-end, delivers encrypted sensitive data, and provides back-end for mini-program back-ends to query for \sessionkey. The vendor refers to the developer of mini-programs that has at least a front-end which resides in the user's social network app, and a back-end server for receiving data sent from the mini-program front-end. Then, the user uses the mini-program and grants permissions for mini-programs to access the sensitive data. \looseness=-1

%In the MK-Centered mini-program resource access, there are mainly three parties: the mini-program developer, the user, and the social network apps. 

\paragraph{Threat Model} We focus on the attacks where attackers can access and unpack the front-end mini-program packages and can capture the network packet of a controlled device (e.g., the attacker's device). 
We assume that the MB is secure (e.g., we cannot compromise the MB by using SQL injection or web shell).  
We also assume that the super app as well as its back-end is secure: the attacker cannot interfere with the super app's execution (such as by hooking the super-app APIs, as the attacker's account will get banned quickly according to our investigation) as well as the super app's back-end.
}

%the cloud services fro free     

\subsection{Scope and Methodologies}
\label{Sub:methodology}

\paragraph{Scope} %In this paper,
  %In this study, we focus specifically on understanding the protection provided by \appsecret-based access protocols in the mini-program ecosystem. 
 In this study, we focus on systematically studying \appsecret leaks to understand their prevalence, root causes, and consequences. We do not aim to detect attacks caused by leakage of the \sessionkey, \token, or \textsf{AT}. This is because as outlined in \S\ref{subsec:threatmodel} and \S\ref{subsec:possibleattacks}, a leaked \appsecret can result in the exposure of \sessionkey and \textsf{AT}, and it is not possible for \token to be leaked. We use the example of \wechat to explain \appsecret-based access control protocols, but note that similar protocols are used on other mini-program platforms, such as \textsf{Baidu}~\cite{baidusmartprogram} and \textsf{Alipay}. Therefore, we focus on \wechat in particular due to its popularity and support for both sensitive data access and cloud service access. 

\paragraph{Collecting Mini-Programs} 
We used the \texttt{innersearch} API (obtained through reverse engineering of \wechat) to search for and download mini-programs, and the \texttt{waVerifyInfo} API to collect developer information~\cite{developinfo21}, if available (as shown in \autoref{tab:inner:apis1}). We employed 1,000 commonly used Chinese characters and 1,000 commonly used English words as seed keywords, expanding them based on the names and descriptions of the collected mini-programs, resulting in 14,020 keywords that retrieved \totalanalyzedapps mini-programs. Note that the \wechat market has about 4 million mini-programs~\cite{wechat3M}, making our dataset likely to cover the majority of them.

 \begin{table}[]
\scriptsize
 
 \setlength\tabcolsep{2pt}\resizebox{0.475\textwidth}{!}{
\begin{tabular}{llcc}\toprule
 {\textbf{API}} &  {\textbf{Description}}    \\
 \midrule

\textbf{{{Mp.wx.qq.com/wxa-cgi/innersearch}}}    &   Mini-program Searching API      \\
   \textbf{Parameters}   &       \\  
\SFviii   \texttt{query}    &   A keyword to search the mini-programs         \\  
%\SFviii Language & \texttt{obj.lang} & \tickNo    &  \tickYes     \\
\SFviii \texttt{client\_version}       &   The \wechat client version    \\ 
\SFviii \texttt{time}       & The timestamp of this query     \\ 
\SFviii  \texttt{cookie} &  \wechat client token     \\
\SFii~ ... & Other omitted parameters\\
\textbf{Return value} &  %The  mini-program metadata  
\\
\SFviii  \texttt{appID}     & The app ID of the mini-program       \\
\SFviii   \texttt{label}     & The category of the mini-program    \\
\SFviii  \texttt{rating}   &  The user rating of the mini-program    \\
\SFii~... & Other omitted fields in the return value\\
%\SFii Profile & A\_A1\_A\_A & a\_a1\_a   &      \\
%~~~\SFii Nick name   & A\_A1\_A\_B & a\_a1\_b   &      \\
  \midrule

\textbf{Wx.nativgateToMiniProgram()}        &    Mini-program Downloading API    \\  
   \textbf{Parameters}   &       \\  
\SFviii\texttt{appID}        &  The app ID of the mini-program          \\  
\SFii\texttt{version}    &  The version (released or debug)       \\ 
\textbf{Return value}    &  The packed file of the mini-program     \\  \midrule

 \textbf{Mp.weixin.qq.com/mp/waVerifyinfo}      & Developer Information Downloading API   \\
    \textbf{Parameters}   &       \\  
\SFviii  \texttt{appID}      &   The app ID of the mini-program     \\
\SFviii \texttt{deviceType}    &  The device type of mobile (e.g., Android)     \\
\SFii\texttt{netType}   &  Network Connection Information (e.g., WiFi) \\
\textbf{Return value}    &     \\  
\SFviii  \texttt{developerInfo}     & The developer information of the mini-program       \\
\SFii\texttt{updateDate} & The relase date of the mini-program\\
\midrule
 \textbf{Api.weixin.qq.com/cgi-bin/token}     & A \appsecret Validation API    \\
    \textbf{Parameters}   &       \\  
\SFviii  \texttt{appID}      &   The app ID of the mini-program     \\
\SFviii \texttt{secret}    &  A \appsecret      \\
\SFii\texttt{grant\_type}   &  The access type  \\
\textbf{Return value}    &   An access token or error message   \\

\midrule

 \textbf{Api.weixin.qq.com/sns/jscode2session}      & \sessionkey query API (\texttt{getEK()})    \\
    \textbf{Parameters}   &       \\  
\SFviii  \texttt{appID}      &   The app ID of the mini-program     \\
\SFviii \texttt{secret}    &  The Master key (\appsecret)      \\
\SFii\texttt{jscode}   &   A login token (\token) \\
\textbf{Return value}    &  The encryption key (\sessionkey)  \\

\bottomrule
\end{tabular} }
 
\caption{The five important APIs used in our study.}
\label{tab:inner:apis1}
\vspace{-0.2in}
\end{table}

\paragraph{Identifying the \appsecret Leaked Mini-Programs} We checked if the mini-programs had leaked their {\tt MK}(s) by identifying all 256-bit hexadecimal digit strings and pruning them based on the \appsecret validation API (the fourth API in \autoref{tab:inner:apis1}). We used a regular expression with a 32-byte hex-string format of \texttt{[a-f0-9]{32}} to search for possible \appsecret{s} in the mini-program code and validated them using the \appsecret validation API. The API returns the actual \textsf{AT} if the \appsecret is correct, else it returns an error code. All mini-programs' backend, by default, has its corresponding \textsf{AT} due to free cloud services like {\tt Translation} as reported in~\autoref{tab:wechatdata}. With the downloaded \totalanalyzedapps mini-programs, we identified \leakedapps mini-programs that leaked their \appsecret{s}.
\looseness=-1

\begin{table*}[]
\footnotesize
 
\setlength\tabcolsep{1pt}
    \centering
    \begin{tabular}{lrrrrrrrrrrrrrrrrrrrrr}
\toprule[1.5pt] 
\multirow{3}{*}{\textbf{Category}}&\multicolumn{12}{c}{{\bf Sensitive Data}}&\multicolumn{9}{c}{\textbf{Cloud Services}}\\ \cmidrule(lr){2-13} \cmidrule(lr){14-22}
% &\multicolumn{3}{c|}{Group Chat Info}&\multicolumn{3}{c}{Shipping Address}\\ \cmidrule{2-22}
&\multicolumn{3}{c}{{\textbf{Phone number (A1)}}}&\multicolumn{3}{c}{{\textbf{User info (A1)}}}&\multicolumn{3}{c}{{\textbf{Share info (A2)}}}&\multicolumn{3}{c}{\textbf{{Werun info (A2)}}}&\multicolumn{3}{c}{{\textbf{AI (A3)}}}&\multicolumn{3}{c}{{\textbf{Security (A3)}}}&\multicolumn{3}{c}{{\textbf{Location (A3)}}}\\\cmidrule(lr){2-4}\cmidrule(lr){5-7} \cmidrule(lr){8-10} \cmidrule(lr){11-13} \cmidrule(lr){14-16} \cmidrule(lr){17-19} \cmidrule(lr){20-22}
% \multicolumn{3}{c|}{\texttt{wx.authPrivateMessage}}&\multicolumn{3}{c|}{\texttt{wx.getEnterGroupInfo}}&\multicolumn{3}{c}{\texttt{wx.chooseAddress}}\\ \cmidrule{2-22}
&\# ind.&\# com.&\% vuln&\# ind.&\# com.&\% vuln&\# ind.&\# com.&\% vuln&\# ind.&\# com.&\% vuln&\# ind.&\# com.&\% vuln&\# ind.&\# com.&\% vuln&\# ind.&\# com.&\% vuln\\ \midrule
Business&0&242&0.59&8&644&1.59&0&49&0.12&0&9&0.02&0&5&0.01&10&855&2.12&0&0&0.00\\
Education&1&775&1.90&114&3,065&7.78&15&635&1.59&0&33&0.08&0&6&0.01&132&3,957&10.00&0&1&0.00\\
e-Learning&0&39&0.10&2&103&0.26&2&22&0.06&0&9&0.02&0&0&0.00&37&134&0.42&0&0&0.00\\
Entertainment&0&42&0.10&6&234&0.59&3&74&0.19&0&2&0.00&0&0&0.00&11&268&0.68&0&0&0.00\\
Finance&0&33&0.08&0&74&0.18&0&10&0.02&0&3&0.01&0&2&0.00&0&96&0.23&0&0&0.00\\
Food&0&511&1.25&8&1,369&3.37&0&158&0.39&0&6&0.01&0&0&0.00&16&1,846&4.55&0&0&0.00\\
Games&1&23&0.06&133&422&1.36&92&170&0.64&0&3&0.01&0&0&0.00&250&473&1.77&0&0&0.00\\
Government&0&141&0.34&24&524&1.34&0&47&0.11&0&7&0.02&0&3&0.01&24&700&1.77&0&0&0.00\\
Health&0&161&0.39&4&448&1.11&1&47&0.12&1&11&0.03&0&4&0.01&4&559&1.38&0&0&0.00\\
Photo&0&2&0.00&1&32&0.08&0&4&0.01&0&1&0.00&0&0&0.00&1&36&0.09&0&0&0.00\\
Job&0&101&0.25&1&275&0.68&0&36&0.09&0&1&0.00&0&2&0.00&2&390&0.96&0&0&0.00\\
Lifestyle&5&1,440&3.53&93&4,183&10.46&4&590&1.45&0&22&0.05&0&9&0.02&124&5,658&14.14&0&1&0.00\\
Shopping&0&1,885&4.61&41&7,174&17.65&6&1,434&3.52&1&65&0.16&0&9&0.02&52&9,504&23.38&0&1&0.00\\
Social&1&55&0.14&13&282&0.72&0&62&0.15&0&6&0.01&0&3&0.01&15&320&0.82&0&0&0.00\\
Sports&1&38&0.10&6&224&0.56&1&25&0.06&3&44&0.11&0&0&0.00&8&253&0.64&0&0&0.00\\
Tool&8&761&1.88&181&3,332&8.59&33&432&1.14&8&43&0.12&0&17&0.04&466&4,327&11.72&2&2&0.01\\
Traffic&1&161&0.40&6&465&1.15&0&33&0.08&0&2&0.00&0&4&0.01&8&665&1.65&0&1&0.00\\
Travelling&0&31&0.08&1&137&0.34&0&12&0.03&0&0&0.00&0&0&0.00&2&191&0.47&0&0&0.00\\
Uncategorized&0&0&0.00&0&3&0.01&0&1&0.00&0&0&0.00&0&0&0.00&0&5&0.01&0&0&0.00\\ \midrule
TOTAL&18&6,441&15.80&642&22,990&57.81&157&3,841&9.78&13&267&0.68&0&64&0.16&1,162&30,237&76.81&2&6&0.02\\

\bottomrule[1.5pt]
 \end{tabular}
 \caption{Statistics of the identified vulnerable mini-programs. Note that ind. stands for individual developers and com. stands for enterprise developers (i.e., by company). %\ZY{Maybe we can color the cells to make them look better?} \ZQ{Explain what does Ind and com mean in the table} 
    }  
    \label{tab:vuln_cat_full}
 
\end{table*}
\begin{table*}[]
    \centering
    \footnotesize
    \setlength\tabcolsep{2.5pt}
 
    \begin{tabular}{lrrrrrrrrrrrrrrrrrr}
    \toprule[1.5pt]
    \multirow{3}{*}{\bf Category}&\multicolumn{18}{c }{\bf Ratings distribution} \\ \cmidrule{2-19}
   &\multicolumn{3}{c}{\textbf{1.0-1.9}}&\multicolumn{3}{c}{\textbf{2.0-2.9}}&\multicolumn{3}{c}{\textbf{3.0-3.9}}&\multicolumn{3}{c}{\textbf{4.0-4.9}}&\multicolumn{3}{c}{\textbf{5.0}}& \multicolumn{3}{c}{\textbf{Unrated}} \\ \cmidrule(lr){2-4} \cmidrule(lr){5-7} \cmidrule(lr){8-10} \cmidrule(lr){11-13} \cmidrule(lr){14-16} \cmidrule(lr){17-19} 
& \# ind.&\# com.&\% vuln&\# ind.&\# com.&\% vuln&\# ind.&\# com.&\% vuln&\# ind.&\# com.&\% vuln&\# ind.&\# com.&\% vuln&\# ind.&\# com.&\% vuln\\ \cmidrule{1-19}
Business&0&0&0.00&0&1&0.00&0&4&0.01&0&38&0.09&0&4&0.01&12&812&2.02\\
Education&0&1&0.00&0&4&0.01&0&38&0.09&7&216&0.55&0&15&0.04&126&3,714&9.39\\
e-Learning&0&0&0.00&0&2&0.00&0&4&0.01&0&11&0.03&0&1&0.00&37&118&0.38\\
Entertainment&0&2&0.00&0&7&0.02&2&17&0.05&1&23&0.06&0&0&0.00&8&226&0.57\\
Finance&0&0&0.00&0&2&0.00&0&2&0.00&0&11&0.03&0&0&0.00&0&82&0.20\\
Food&0&0&0.00&0&1&0.00&0&2&0.00&0&49&0.12&0&2&0.00&16&1,795&4.43\\
Games&0&0&0.00&18&23&0.10&63&124&0.46&9&37&0.11&0&0&0.00&161&294&1.11\\
Government&0&0&0.00&0&2&0.00&0&7&0.02&0&48&0.12&0&1&0.00&24&645&1.64\\
Health&0&0&0.00&0&1&0.00&0&3&0.01&0&49&0.12&0&2&0.00&4&510&1.26\\
Photo&0&1&0.00&0&0&0.00&0&2&0.00&1&7&0.02&0&0&0.00&0&27&0.07\\
Job&0&0&0.00&0&1&0.00&0&10&0.02&0&29&0.07&0&1&0.00&2&351&0.86\\
Lifestyle&0&0&0.00&0&1&0.00&0&37&0.09&3&227&0.56&1&19&0.05&120&5,398&13.50\\
Shopping&0&2&0.00&0&0&0.00&2&23&0.06&10&299&0.76&0&26&0.06&42&9,205&22.62\\
Social&0&0&0.00&0&3&0.01&0&5&0.01&1&38&0.10&0&0&0.00&14&279&0.72\\
Sports&0&0&0.00&0&0&0.00&0&2&0.00&0&33&0.08&1&6&0.02&7&213&0.54\\
Tool&0&3&0.01&3&24&0.07&4&48&0.13&19&325&0.84&4&20&0.06&441&3,955&10.75\\
Traffic&0&0&0.00&0&5&0.01&0&16&0.04&0&71&0.17&0&3&0.01&8&574&1.42\\
Travelling&0&0&0.00&0&0&0.00&0&3&0.01&0&12&0.03&0&1&0.00&2&175&0.43\\
Uncategorized&0&0&0.00&0&0&0.00&0&0&0.00&0&0&0.00&0&0&0.00&0&5&0.01\\ \midrule
TOTAL&0&9&0.02&21&77&0.24&71&347&1.02&51&1,523&3.85&6&101&0.26&1,024&28,378&71.92\\
 
\bottomrule[1.5pt]
% business&0&0&0&0&0&0&1&3&3&37&0&570&614\\
% education&0&0&0&2&1&4&8&34&75&145&0&3,290&3,559\\
% entertainment&0&0&1&1&2&6&9&10&14&16&0&188&247\\
% finance&0&0&0&0&0&1&0&2&6&4&0&59&72\\
% food&0&0&0&0&1&0&1&1&13&37&0&1,523&1,576\\
% games&0&0&1&3&23&68&135&182&51&12&0&1,252&1,727\\
% government&0&0&0&0&0&0&1&5&17&32&0&556&611\\
% health&0&0&0&0&1&0&1&4&4&36&0&392&438\\
% images&0&0&0&1&0&0&1&1&2&6&0&25&36\\
% job&0&0&0&0&0&1&4&6&14&13&0&280&318\\
% lifestyle&0&0&0&0&0&0&7&24&81&167&0&4,517&4,796\\
% reading&0&0&0&0&1&1&1&2&5&5&0&177&192\\
% shopping&0&0&0&1&1&0&5&20&95&239&0&6,887&7,248\\
% social&0&0&0&0&2&2&2&3&20&13&0&274&316\\
% sports&0&0&0&0&0&0&0&1&5&31&0&174&211\\
% tool&0&0&1&3&14&16&26&39&115&246&0&4,665&5,125\\
% traffic&0&0&0&0&1&4&4&13&26&50&0&449&547\\
% travelling&0&0&0&0&0&0&0&1&1&10&0&124&136\\
% uncategorized&0&0&0&0&0&0&0&0&0&0&0&3&3\\
    \end{tabular}
    \caption{Distribution of the detailed ratings of mini-programs} 
    \label{tab:vuln_ratings} 
   
\end{table*}

\looseness=-1

\subsection{The Results}
\ignore{
In this section, we present the evaluation results. Particularly, we seek to obtain the insights from the vulnerable mini-programs by answering the following research questions:

\begin{itemize}
\item \textbf{RQ1:} What are the categories of vulnerable mini-programs?
\item \textbf{RQ2:} What are the ratings of the vulnerable mini-programs?
\item \textbf{RQ3:} What are the accessed resources of vulnerable mini-programs?
\item \textbf{RQ4:} Who are the developers of vulnerable mini-programs?
\item \textbf{RQ5:} What are the lastest update of the vulnerable mini-programs?
\item \textbf{RQ6:} Are there any high profile vulnerable mini-programs?

\end{itemize}
}

With the detected \leakedapps mini-programs, we next seek to answer the research questions set out earlier by analyzing their category, ratings, accessed resources, their developers, and last update. %(i) their category it belongs to; (ii) the rating of the vulnerable mini-programs, (iii) the privacy resources it accesses (e.g., \texttt{weRunData}, \texttt{phoneNumber}); (iv) when the miniapp was last updated;  and (v) the difference between enterprise developers and individual developers. 
In the following, we provide these results in greater details.  

%\begin{enumerate} [label=(\Roman*)]

\paragraph{(RQ1) Categories of the \appsecret Leaked Mini-programs} Most mini-programs in the \wechat mini-program store have been classified into specific categories, as shown in \autoref{tab:vuln_cat_full}. We observed that the percentage of \appsecret leaked mini-programs and the data or services they accessed (e.g., phone number, user info, share info, werun info, AI, security, and location) was diverse, indicating that \appsecret leakage was widespread across mini-program categories. However, some categories (e.g., lifestyles and shopping) had a slightly higher number of vulnerable mini-programs due to the larger number of mini-programs in those categories. On the other hand, tool, government, and games had a higher percentage of vulnerable mini-programs. Unfortunately, all these mini-programs are crucial for users' daily lives as they often provide essential services (e.g., QR code scanning), involve real user identity information and critical administrative services (e.g., government tools), or involve payment (e.g., shopping) and in-game trading.\looseness=-1

\begin{figure*}
    \centering
        \includegraphics[width=.245\linewidth]{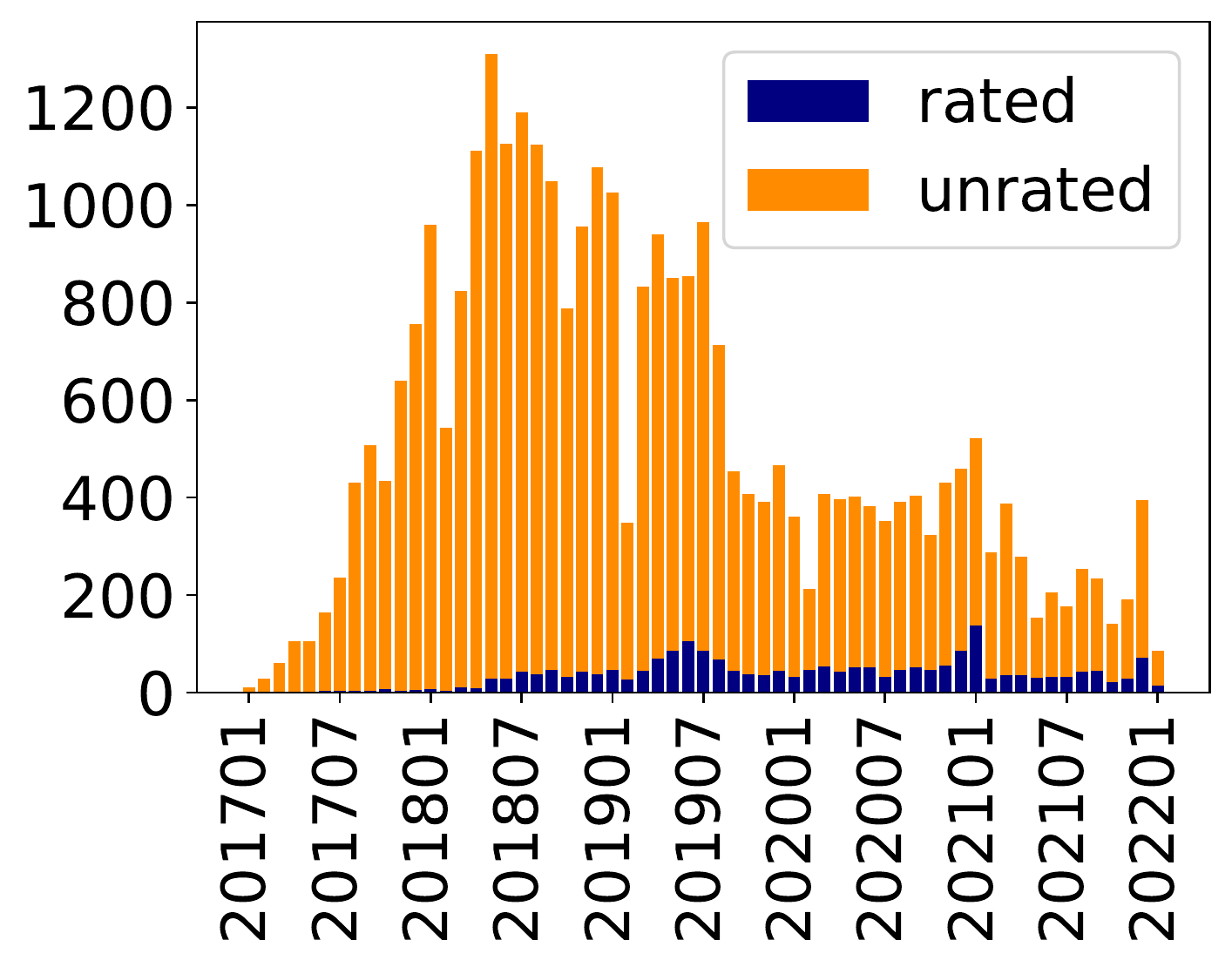}
        \includegraphics[width=.245\linewidth]{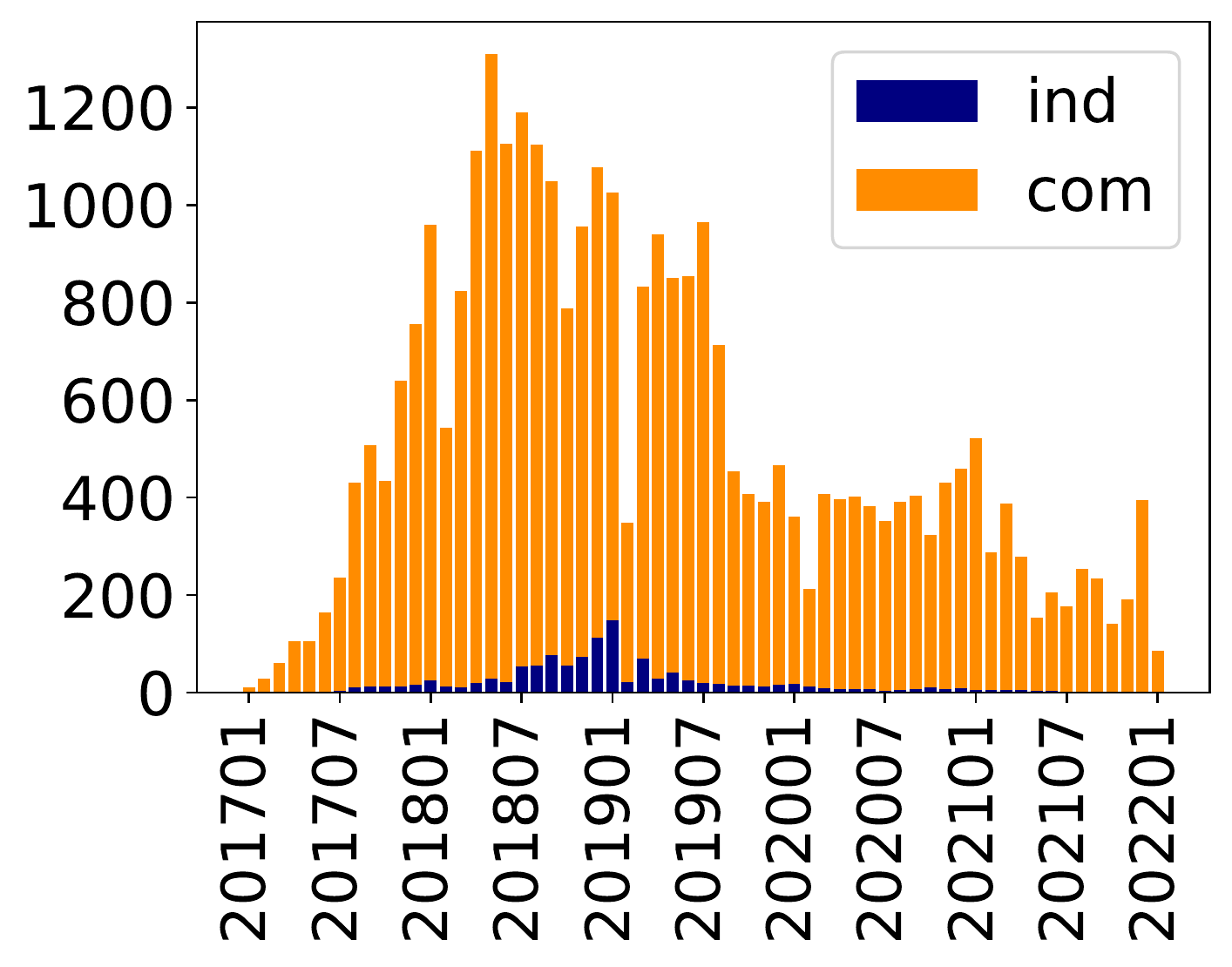}
        \includegraphics[width=.245\linewidth]{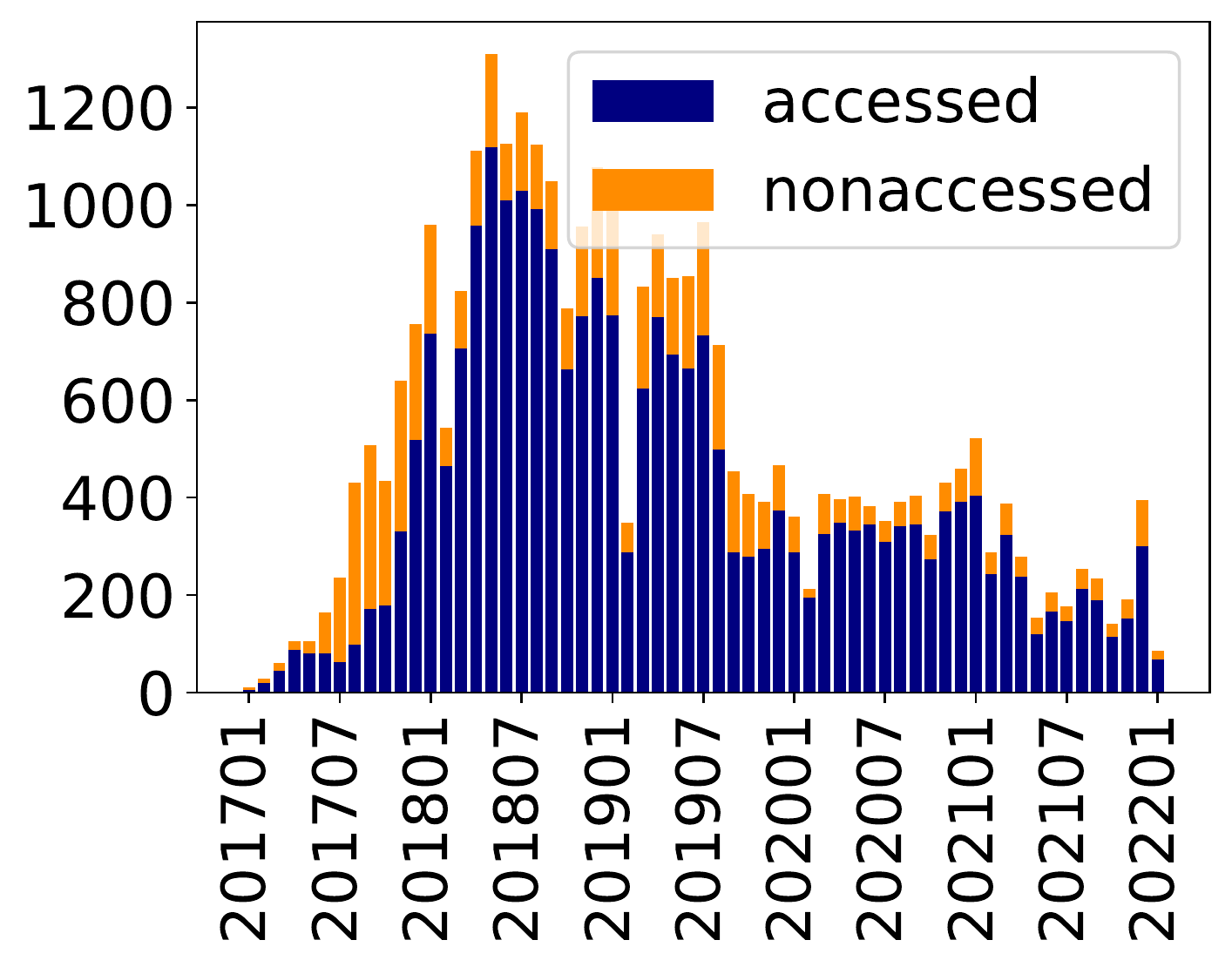}
        \includegraphics[width=.245\linewidth]{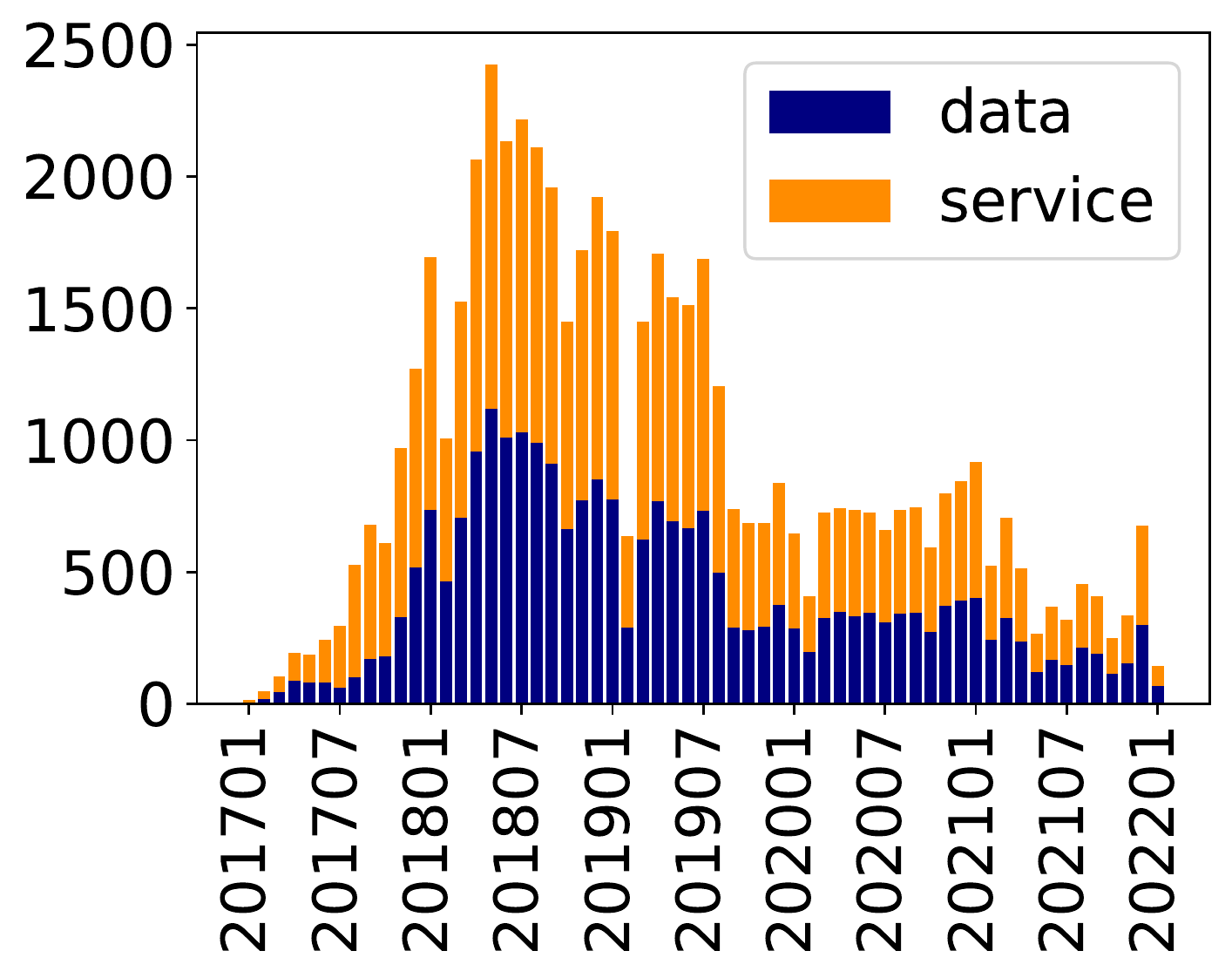}
     
    \caption{The trending of all vulnerable mini-programs w.r.t rating, developer, accessed resource type, and accessed data type.}
    \label{fig:trends}
    
\end{figure*}

%\vspace{-0.2cm} 
\paragraph{(RQ2) Ratings of the \appsecret Leaked Mini-programs} Compared to traditional platforms like Google Play, the mini-program store does not furnish information about the total number of downloads for each mini-program. Nevertheless, we can estimate its popularity by evaluating its rating. It is worth noting that the rating becomes available only after a sufficient number of users have employed and rated the corresponding mini-program~\cite{ourfancywork}. By utilizing this rating, we can gauge the popularity of a mini-program and link it to the corresponding mini-program category, as demonstrated in \autoref{tab:vuln_ratings}. The table illustrates that mini-programs with lower ratings are more likely to have a greater number and percentage of vulnerable mini-programs (i.e., \appsecret leaked mini-programs). This is likely because developers of mini-programs with higher ratings may devote more effort to ensuring security and privacy to maintain their competitive edge. \looseness=-1

\paragraph{(RQ3) Accessed Resources of the \appsecret Leaked Mini-programs}  
As shown in \S\ref{subsec:attackworkflow}, attackers can use different attacks depending on the sensitive resources accessed by mini-programs. To understand the impact of \appsecret leaks in these vulnerable mini-programs, we present statistics in \autoref{tab:vuln_cat_full} based on APIs used. In particular, we examined the API names directly in the code to identify mini-programs that utilized sensitive data. Moreover, we employed \texttt{api.weixin.qq.com/cgi-bin/get\_current\_selfmenu\_info} to inspect the specific services utilized by a mini-program (without invoking these services to verify their availability).  Our findings show that \numaccesseddata mini-programs accessed sensitive data and \numaccessedcloud accessed at least one cloud service, making them vulnerable to attacks such as account hijacking (A1), promotion abuse (A2), and service theft (A3) (The details of those attacks will be introduced in \S\ref{subsec:attackworkflow}). Interestingly, we also discovered that there are \nonaccessed vulnerable mini-programs that do not access any MK-related resources at all. However, these mini-programs may still be vulnerable since they could potentially be utilized to access MK-related resources in the future.
%As to be demonstrated in \S\ref{subsec:attackworkflow}, an attacker can launch different types of attacks depending on the sensitive resources a mini-program accesses. To understand the potential impacts of \appsecret leakage among these vulnerable mini-programs, we present statistics on the different sensitive data accessed by these mini-programs in \autoref{tab:vuln_cat_full} based on the APIs used. We found that \numaccesseddata mini-programs have accessed various sensitive data, including phone numbers, werun data, user information, and mini-program share information, and \numaccessedcloud mini-programs have accessed at least one cloud service. All of these mini-programs that access data or service resources are vulnerable to a variety of attacks, including account hijacking (A1), promotion abuse (A2), and service theft (A3). Interestingly, we also found that there are \nonaccessed vulnerable mini-programs that do not access any MK-related resources at all. However, these mini-programs could still be vulnerable, since they could potentially be used to access MK-related resources in the future.

\begin{table}[]
\footnotesize
 \setlength\tabcolsep{3pt}
{ 
\begin{tabular}{lrrrrrr}
\toprule[1.5pt]
\multirow{2}{*}{\textbf{\begin{tabular}[c]{@{}c@{}}\# Associated\\  Apps\end{tabular}}}                   & \multicolumn{2}{c}{\textbf{Rated}}             & \multicolumn{2}{c}{\textbf{Unrated}}           & \multicolumn{2}{c}{\textbf{Total}}             \\ \cmidrule(lr){2-3} \cmidrule(lr){4-5} \cmidrule(lr){6-7}
 & \textbf{\# com} & \textbf{\# apps} & \textbf{\# com} & \textbf{\# apps} & \textbf{\# com} & \textbf{\# apps} \\ \hline
>=20&0&0&16&417&16&417\\
19&0&0&2&38&2&38\\
18&0&0&2&36&2&36\\
17&0&0&1&17&1&17\\
16&0&0&3&48&3&48\\
15&1&15&2&30&3&45\\
14&0&0&1&14&1&14\\
13&1&13&4&52&5&65\\
12&3&36&6&72&9&108\\
11&0&0&1&11&1&11\\
10&0&0&9&90&9&90\\
9&3&27&9&81&12&108\\
8&1&8&24&192&25&200\\
7&1&7&29&203&30&210\\
6&1&6&26&156&27&162\\
5&6&30&69&345&75&375\\
4&7&28&155&620&161&648\\
3&21&63&333&999&353&1,062\\
2&100&200&1,300&2,600&1,388&2,800\\
1&1,624&1,624&22,357&22,357&23,805&23,981\\ \midrule
TOTAL&1,769&2,057&24,349&28,378&25,781&30,435\\
 \bottomrule[1.5pt]

\end{tabular} 
}
    \caption{Numbers of vulnerable mini-programs associated with the same software development enterprises.} 
    \label{tab:company}
  
\end{table}

\paragraph{(RQ4) The Developers of the \appsecret Leaked Mini-programs} We now investigate the developers of these vulnerable mini-programs and their frequency of errors. There are two types of developers: individual and enterprise. \Tencent provides enterprise name and tax ID for enterprise developers, but not for individual developers. Thus, we can only present statistics for vulnerable mini-programs developed by companies, excluding 4,676 developed by individuals. For companies, we collected developer information if available, and \autoref{tab:company} shows the statistics of associated mini-programs. Most companies developed only one vulnerable mini-program, but some, such as ``Suzhou Yutianxia'', developed 30.

% Next, we aim to understand who are the developers of these vulnerable mini-programs, 
% and how often they made these mistakes. 
% At a high level, there are two kinds of developers: individual developer and enterprise developers. 
%  Compared with enterprises developers whose enterprise name and tax ID is shown in the metadata, \Tencent does not provide such info for mini-programs developed by individual developers, whereas the metadata page only shows `individual developer'.  
% Therefore, in the following, other than 4,676 mini-programs that have been developed by the individual developers, we present the 
% statistics of the vulnerable mini-programs that are produced by the corresponding companies.  
% In particular, we have collected the developer information of the companies of the vulnerable mini-programs if they are available, and \autoref{tab:company} shows the statistics based on how many mini-programs have been associated with the same company. We can notice that the majority of the companies developed just a single vulnerable mini-program, and but sometimes one company can produce multiple vulnerable mini-programs (e.g., a company called `Suzhou Yutianxia' has produced 30 vulnerable mini-programs according to the data we have). 

\begin{table}[t]
  \centering
    \footnotesize
     
\begin{tabular}{@{}lcllr@{}}
\toprule[1.5pt]
\textbf{Company}          & \textbf{Ratings}      & \textbf{Category} & \textbf{Data}                    & \textbf{Attacks} \\ \midrule
\multirow{17}{*}{Tencent}      & 5.0                  & games             & \ding{192}                       & A1               \\
                              & 5.0               & tool              & \ding{192}                       & A1               \\
                              & 5.0                      & entertainment     & \ding{192},\ding{193}            & A1, A2           \\
                              & 4.6            & health            &                                  &                  \\
                              & 5.0                   & entertainment     & \ding{192}                       & A1               \\
                              & 4.8                     & tool              & \ding{192}                       & A1               \\
                              & 4.7               & education         &                                  &                  \\
                              & 4.6                  & games             & \ding{192}                       & A1               \\
                              & 4.7                 & social            & \ding{192}                       & A1               \\
                              & 4.6       & tool              & \ding{192}                       & A1               \\
                              & 5.0                   & tool              &                                  &                  \\
                              & 5.0     & e-learning        & \ding{194}                       & A2               \\
                              & 5.0                       & health            & \ding{192}                       & A1               \\
                              & 5.0                       & games             & \ding{192}                       & A1               \\
                              &                & games             & \ding{192}                       & A1               \\
                              &                             & education         &                                  &       A3           \\
                              & 5.0                             & games             & \ding{193}                       & A2               \\ \midrule
\multirow{6}{*}{Nestle}        & 4.3                      & food              & \ding{194},\ding{192}            & A1, A2           \\
                              & 5.0           & tool              & \ding{194},\ding{192}            & A1, A2           \\
                              &              & lifestyle         & \ding{194},\ding{192}            & A1, A2           \\
                              & 4.9            & food              & \ding{194}                       & A2               \\
                              & 4.8            & shopping          & \ding{194},\ding{192}            & A1, A2           \\
                              & 4.9                    & shopping          & \ding{194},\ding{192}            & A1, A2           \\ \midrule
\multirow{4}{*}{Sanofi}        &  4.7               & health            & \ding{194},\ding{192}            & A1, A2           \\
                              & 5.0                    & business          & \ding{194},\ding{192}            & A1, A2           \\
                              & 5.0                   & business          & \ding{194},\ding{192}            & A1, A2           \\
                              &                   & business          & \ding{194},\ding{192}            & A1, A2           \\ \midrule
\multirow{4}{*}{China Unicom}  & 5.0        & finance           & \ding{194}                       & A2               \\
                              & 5.0                       & finance           & \ding{194},\ding{192}            & A1, A2           \\
                              &  4.8               & tool              &                                  &                  \\
                              & 4.1            & shopping          & \ding{194}                       & A2               \\ \midrule
\multirow{3}{*}{Bank Of China} & 4.3                    & shopping          & \ding{192}                       & A1               \\
                              & 4.3                & shopping          & \ding{192}                       & A1               \\
                              &                  & business          & \ding{192}                       & A1               \\\midrule
\multirow{3}{*}{Hitachi}       & 5.0                  & tool              & \ding{192}                       & A1               \\
                              & 5.0                     & tool              & \ding{192}                       & A1               \\
                              & 5.0                       & business          & \ding{192}                       & A1               \\\midrule
\multirow{2}{*}{Shou Gang}     &             & tool              &                                  &        A3          \\
                              & 5.0  & lifestyle         &                                  &                  \\
\multirow{2}{*}{China Mobile}  & 4.3            & finance           &                                  &                  \\
                              & 4.7             & finance           &                                  &                  \\ \midrule
\multirow{2}{*}{ICBC}          & 5.0          & lifestyle         & \ding{194}                       & A2               \\
                              &               & traffic           & \ding{192},\ding{193},\ding{195} &                  \\ \midrule
\multirow{2}{*}{Volkswagen}    & 5.0              & shopping          & \ding{192}                       & A1               \\
                              &              & shopping          & \ding{192},\ding{193}            &                  \\ \bottomrule[1.5pt]
\end{tabular}
\caption{The top-10 Fortune 500 companies that have the most number of mini-programs leaked the \appsecret. \ding{192} user info; \ding{193} shared info; \ding{194} phone number; \ding{195} werun data. 
%A1: Account hijack attacks; A2: promotion abuse attacks.  A3: Service theft attack.
    }
    \label{tab:appsecret}
 
\end{table} 

\paragraph{(RQ5) Latest Update of the \appsecret Leaked Mini-programs} The mini-program's metadata includes the latest updated timestamp, allowing us to determine how long the \appsecret leakage vulnerability has existed. We also analyze the resources accessed, ratings, and developer information grouped by the latest update month to observe trends over time (\autoref{fig:trends}). Our observations are as follows: (i) \appsecret leakage vulnerability has been there since the first year when \Tencent introduced the mini-program into \wechat (e.g., a Calendar mini-program leaked the \appsecret in January 2017 and was never updated). (ii) Over time, the number of vulnerable mini-programs decreased regardless of the accessed data resource or rating, but those developed by companies still constitute a significant portion of the vulnerable mini-programs.
% decreased significantly.
% \AY{Interestingly, we were also surprised to find out that although the problem of \appsecret leakage had been raised by developer as early as in 2019~\cite{}, the \appsecret leakage continues to exists even after 2021. This further demonstrated the surprising neglect of potential impacts brought about by \appsecret leakage among the mini-program community.}

%\end{enumerate}

\paragraph{(RQ6) High Profile \appsecret Leaked Mini-programs} Lastly, we investigate the implications of our attacks by identifying the vulnerable mini-programs. We found \bigcompanies mini-programs developed by Fortune top-500 companies, and those with multiple vulnerable mini-programs are listed in \autoref{tab:appsecret}. Notably, (i) some high-profile companies made the same mistake across multiple mini-programs (e.g., Nestle); (ii) Tencent, the \wechat provider, is the top-ranked company with \appsecret leaks in their mini-programs; (iii) some mini-programs from high-profile companies only leaked their \appsecret without accessing sensitive information. As discussed, while these mini-programs cannot be attacked currently, they remain vulnerable if new functionalities are added (\eg, phone number fetching).

\ignore{
To gain a better understanding of the \appsecret exposure, we have conducted analysis over the \leakedapps vulnerable miniapps, through which we have  further obtained more insights from them such as (I) the distribution 
of the ratings of the vulnerable miniapps
(II) distribution of the categories of the vulnerable miniapps,  (III) the distribution of the sensitive information these vulnerable miniapps accessed, (IV) the distribution of the developer information of these vulnerable miniapps, and (V) the distribution of vulnerable miniapp registration and updating date. 
}

\section{Exploiting the \appsecret Leaks}
\label{subsec:attackworkflow}

We have found vulnerable mini-programs that exposed their app secrets, allowing for two types of attacks: (i) attacks against sensitive data (\S\ref{attack:data}), where the attacker aims to modify sensitive data; and (ii) attacks against cloud services (\S\ref{attack:services}), where the attacker uses the victim mini-program's paid cloud services without cost.

\subsection{Attacks Against Sensitive Data}
\label{attack:data}

\begin{figure}
    \centering
    
     \includegraphics[width=\linewidth]{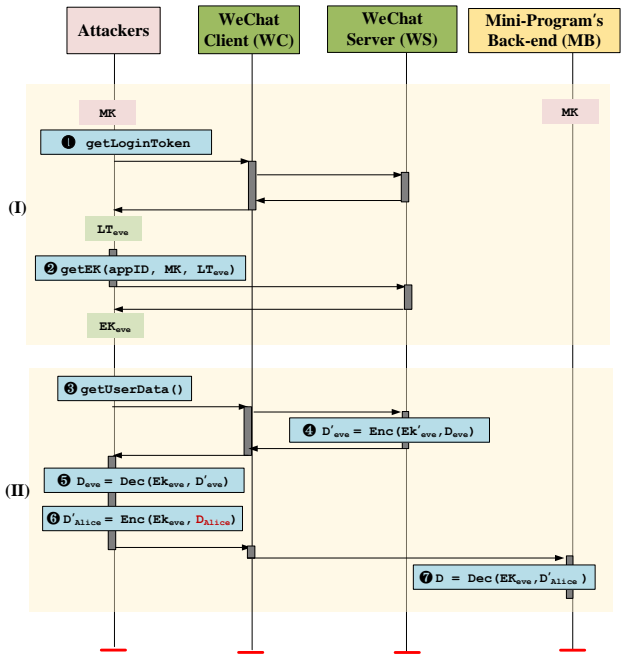}
     \vspace{-0.25in}
    \caption{Attacks Against Sensitive Data}
    \label{fig:attack}
   % \vspace{-0.1in}
\end{figure}

\paragraph{Attack Workflow} 
In our threat model, the attacker can obtain the mini-program front-end package and manipulate network packets. This allows them to unpack the program and trigger sensitive data communication for retrieval or manipulation. The attack workflow involves two steps as shown in \autoref{fig:attack}: \looseness=-1

\begin{enumerate}[label=(\Roman*)]
 
\item \textbf{Obtaining Attacker's Encryption Key (\sessionkey).} To start the attack, assume an attacker ${Eve}$, he needs to first get his own \token, denoted {\tt LT}$_{eve}$, 
from the \wechat server (\textbf{Step \ding{182}}). Next, he provides the leaked \appsecret, the mini-program {\tt appID} (easily obtained in the meta-data of a mini-program), and {\tt LT}$_{eve}$ to the \wechat server (\textbf{Step \ding{183}}). Since \appsecret,  {\tt appID}, and {\tt LT}$_{eve}$ are all valid, the \wechat server will send his \sessionkey (\ie, {\tt EK}$_{eve}$) as what it does typically for the mini-program's back-end to ${Eve}$. Note that prior to this login, the server may already have a copy of {\tt EK}$_{eve}$ if $Eve$ has logged in before the mini-program server. Also, if \wechat server has refreshed the {\tt EK}$_{eve}$,  ${Eve}$ will get a new one, and the mini-program back-end will detect the {\tt EK} change as well  (at \textbf{Step \ding{187}}) and request for this new {\tt EK}$_{eve}$ from the \wechat server by asking $Eve$ to provide his $LT_{eve}$ again (details are omitted here for brevity). Therefore, it does not matter whether the mini-program server has a copy of {\tt EK}$_{eve}$ or not. \looseness=-1

\item \textbf{Sensitive Data Retrieval and/or Manipulation.} In this stage, the attacker ${Eve}$ initiates a request for sensitive data (\eg, \texttt{getPhoneNumber}) and prompts \wechat client to retrieve his or her own data (Step \ding{184}). The \wechat client sends a request to its server, which encrypts the requested data such as $Eve$'s Phone number $D_{eve}$ (Step \ding{185}). The server returns the encrypted data, represented by $D_{eve}^\prime$, to ${Eve}$, who can then choose to discard it and manipulate the information. Nevertheless, the attacker must first decrypt the cipher to examine the packet format since they are unaware of the data's structure (Step \ding{186}). Only then can the attacker create falsified data.  Next, the attacker re-encrypts the fake data using the same \sessionkey (\textbf{Step \ding{187}}).
For example, ${Eve}$ can change his phone number to Alice's (\ie, $D_{Alice}$). 
%decrypted data depending on his or her goal. For example, the attacker can change his own phone number to someone else's (\textbf{Step} \ding{201}). Next, 
%The attacker then sends the encrypted data back to the mini-program's backend (\textbf{Step \ding{186}}). 
Since the mini-program's backend has the same \sessionkey as the attacker, it can decrypt the cipher and obtain the modified data (\ie, phone number $D_{Alice}$) (\textbf{Step \ding{188}}). If there is no consistency check between the phone number and the \sessionkey (likely true at this moment according to our experiment), the attacker can break into Alice's account. \looseness=-1
\end{enumerate}

Now, the attacker can directly obtain and modify the information of interests. It turns out such data manipulation can cause severe consequences. % However, the attacker can only obtain and modify his or her own sensitive data.  what are the consequences of such attacks? 
In the following, we would like to demonstrate the implication of these attacks by using two concrete examples unique in the super apps: account hijacking   and promotion abuse.   

\begin{figure*}[t]
    \centering
    \includegraphics[width=0.85\linewidth]{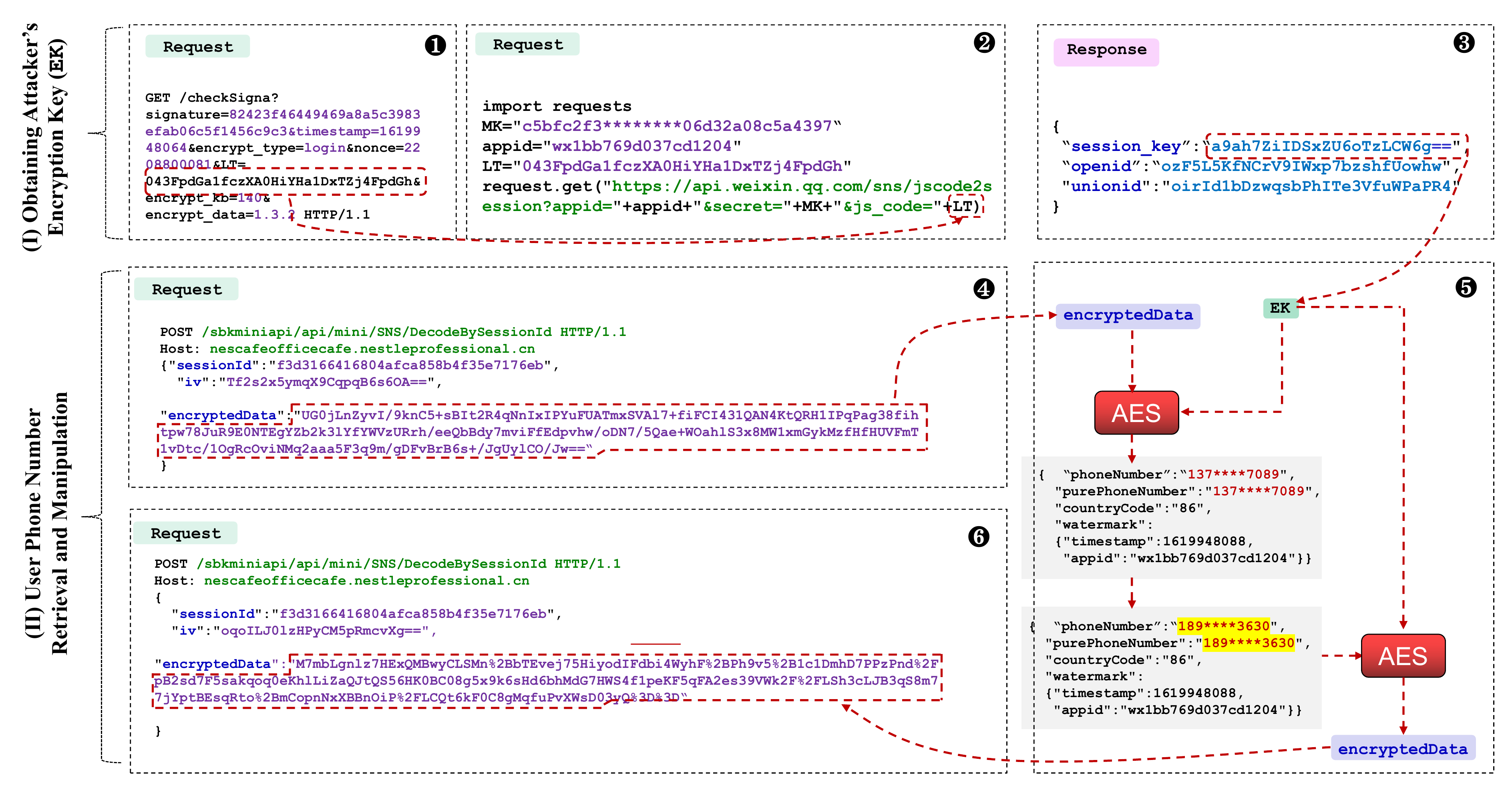} 
    \vspace{-0.15in}
    \caption{An excerpt of attack traffic traces  }
    \label{fig:attackexample}
%\vspace{-0.2in}
\end{figure*}

\paragraph{(A1) Account Hijacking Attacks} 
Since a mini-program's back-end usually needs to maintain a database to keep user's record and this database needs to be indexed, the mini-program back-ends usually use predictable data such as the phone number to index each user (this is particularly true in China since most citizen has a cell phone number, and this number has been used almost as an identity). As such, with \appsecret, the attacker can retrieve the victim's information by changing the predictable data (\eg, the phone number) to a victim's, logging into other users' accounts for illicit purposes. For instance, he or she can steal the discount offers or even free offers given to high-value loyal customers, such as hotel free night certificates or rewards points. % especially for some hotels that give high-value customers large discounts. The attacker may be able to redeem for a free night for a luxury hotel if the victim has such an offer in his or her account. 

%While these attacks can be targeted (i.e., the attacker knows the phone numbers of the victim user before launching the attacks), it is also possible that an attacker can enumerate the phone number to launch the attack at scale. What makes it even worse is that the limited enumeration space of these predictable data that make the attack easier.
% the attacker may be able to get the sensitive information of victim's address, identity information such as name and ID, and order information that may contain home address. Besides, financial losses may be caused to the user or vendor if the attacker uses discount offers given to the victims (especially when some hotels and shops give high-value users large discounts or even free offers). The privacy leakage may also be conbined with other
%For example, the 10-digit phone number consists of 3-digit area code, 3-digit central office code, and 4 digits of another enumerable code. As such, the enumeration space of phone number is quite small.  

%\paragraph{Account Hijacking via \texttt{PhoneNumber} Manipulation}

We now present such a concrete attack  % via phone number manipulation 
against a mini-program named ``{\tt We Proudly Serve}'' (WPS) from \textit{Nestle} to demonstrate its consequences. Note that this mini-program has fixed the vulnerability after we made the responsible disclosure. %real impacts. %Before our attack, we first unpack the front-end of \texttt{StarBucks} mini-program using \texttt{WXUnpacker}~\cite{decomplier}, and extract \appsecret. We save the \appsecret for further usage.
Specifically, to carry out our attack, we first registered two accounts, a victim account and an attack account, and then executed the following steps: %, as we must make sure we attack our own accounts, a community practice for the attack works.
\looseness=-1
\begin{enumerate} [label=(\Roman*)]
 
\item \textbf{Obtaining Attacker’s Encryption Key (\sessionkey).} In this stage, we first got \token from \wechat server, and used the obtained \token, mini-program {\tt appID}, and the leaked \appsecret to query the encryption key \sessionkey (\textbf{Step \ding{182}}).  To get the \token from \wechat server, we used a well-known MITM proxy \texttt{Burp Suite}~\cite{wear2018burp} to inspect the traffic between the \wechat client and its server.  By default, when a victim mini-program invokes \texttt{wx.login},  the \token will be delivered to the front-end and require the front-end to send it to the back-end. In our attack, we intercepted the \token (i.e., the \texttt{encrypt\_code} in \autoref{fig:attackexample}) without sending it to the back-end as shown in  \autoref{fig:attackexample}.  Next, we programmed a python script to query the \sessionkey from \wechat server, and the obtained \token was then fed to the \texttt{getEK()} API (\textbf{Step \ding{183}}). Specifically, \texttt{getEK()} takes three inputs, including the public accessible appID,  the \texttt{secret} (\ie, \appsecret), and  \texttt{Js\_code} (\ie, the \token obtained in \textbf{Step \ding{182}}). As shown in \autoref{fig:attackexample}, \wechat server returned the encryption key \sessionkey, which is encoded using Base64 (\textbf{Step \ding{184}}). \looseness=-1

\item \paragraph{Phone Number Retrieval and Manipulation} In this stage,  we intentionally triggered  WPS's front-end to invoke \texttt{getPhoneNumber}, and manipulated the encrypted phone number sent from the \wechat server to trick the back-end of WPS into believing it is interacting with the victim's account.
More specifically, we first triggered  WPS to invoke \texttt{getPhoneNumber}. 
\texttt{getPhoneNumber} is a callback function, which is called when we click the corresponding button.
Then we intercepted the cipher that is about to be sent to the mini-program back-end (\textbf{Step \ding{185}}), and this cipher was initially received by the mini-program from \wechat server when invoking the \texttt{getPhoneNumber} API. 
According to \wechat's official document~\cite{wechatkey20}, \wechat uses the standard cryptographic algorithms, and therefore, we easily implemented our own script to decrypt and encrypt the {\tt encryptedData}
using the obtained \sessionkey,  and inspected its format, so that we can forge a fake phone number in the following steps (\textbf{Step \ding{186}}). After inspecting the format of the decrypted plain-text, we then replaced the attacker's phone number  (\ie,\breakabletexttt{137****7089}) with our victim's (\ie, \breakabletexttt{189****3630"}), and encrypted the modified phone number using \sessionkey. At this time, the attacker can send the encrypted data (the plaintext of which is the modified phone number) to the back-end of WPS (\textbf{Step} \ding{187}).  
As discussed, the \sessionkey used by the backend of WPS to decrypt the phone number and \sessionkey used by the attacker to encrypt the phone number are the same,  and therefore, our modified phone number was successfully decrypted by the back-end of WPS, and used by WPS to retrieve the user information.  It is interesting to note that much of the retrieved information is collected and maintained by WPS, meaning that if we hijack multiple accounts by enumerating all the phone numbers, we can harvest a large amount of WPS's user information. 
\end{enumerate}

Even with root access to a phone, attackers cannot manipulate encrypted data without the MK. While plaintext phone numbers can be changed during user registration, they are protected by SMS authentication. Later on, all phone numbers are encrypted and sent directly to the WeChat server, never exposing plaintext to the front-end of the mini-program or WeChat app. Without MK, attackers won’t be able to mount the account hijacking. \looseness=-1
 
\begin{figure} [h]
\centering
 \centering
    \includegraphics[width=\linewidth]{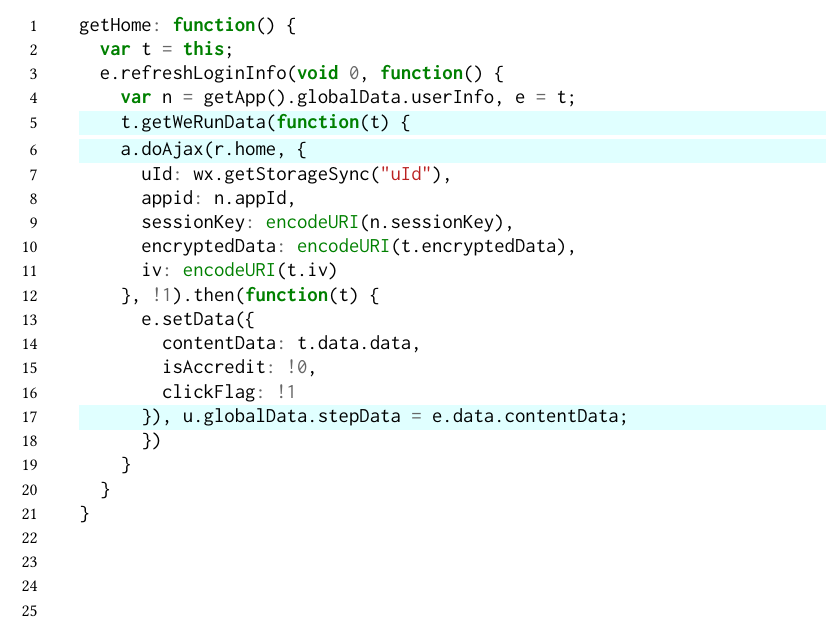} 
% \begin{minted}[linenos,fontsize=\scriptsize,highlightlines={5, 6, 17}, xleftmargin=.05\textwidth]{Javascript}
% getHome: function() {
%   var t = this;
%   e.refreshLoginInfo(void 0, function() {
%     var n = getApp().globalData.userInfo, e = t;
%     t.getWeRunData(function(t) {
%     a.doAjax(r.home, {
%       uId: wx.getStorageSync("uId"),
%       appid: n.appId,
%       sessionKey: encodeURI(n.sessionKey),
%       encryptedData: encodeURI(t.encryptedData),
%       iv: encodeURI(t.iv)
%     }, !1).then(function(t) {
%       e.setData({
%         contentData: t.data.data,
%         isAccredit: !0,
%         clickFlag: !1
%       }), u.globalData.stepData = e.data.contentData;
%       })
%     }
%   }
% }

% \end{minted}

     \caption{A vulnerable mini-program from ICBC}
          \label{list:pinyue}
      
\end{figure}

\paragraph{(A2) Promotion Abuse Attacks} 
Attackers can manipulate sensitive data other than phone numbers, such as {\tt WeRunData} and {\tt ShareInfo}. In the mini-program paradigm, promoting services to more users is crucial for success, and convenient access to social networks empowers mini-programs to promote products more effectively. Some mini-programs offer incentives, such as sharing promotions via group chats, which can range from trial products to real cash. However, attackers who manipulate encrypted {\tt WeRunData} or {\tt ShareInfo} can easily exploit these promotions and earn rewards without actually sharing promotion information to multiple group chats by manipulating group chat identifiers.
\paragraph{Case I: Promotion Abuse via \texttt{WeRunData} Manipulation} 
Promotion abuse attacks can be achieved through manipulating \texttt{WeRunData} data. For example,  
we discovered that a banking mini-program  from Industrial and Commercial Bank of China (ICBC) that uses the \texttt{WeRunData} for promotion where users gain the ``energetic value'' with the steps they walked each day, and such a value can be used to redeem for gifts in the shop. By inspecting the unpacked JavaScript code (which can be found in \autoref{list:pinyue}), we found that \texttt{WeRunData} is obtained with \texttt{getWeRunData}  and then the encrypted data is sent to the back-end via AJAX request. % together with the \sessionkey. 
However, as the \appsecret is leaked, the attacker can abuse these promotions by arbitrarily setting the daily steps walked across a week and sending the request to the mini-program back-end. Specifically, the attacker can first use \sessionkey to decrypt \texttt{t.encryptedData} and change the original daily steps from, for example, zero to a hundred thousand. Given that the attacker can have multiple accounts, theoretically,  the attacker can perform the attack at a scale to harm the vendors. \looseness=-1

 \begin{figure} 
\centering
 \centering
    \includegraphics[width=\linewidth]{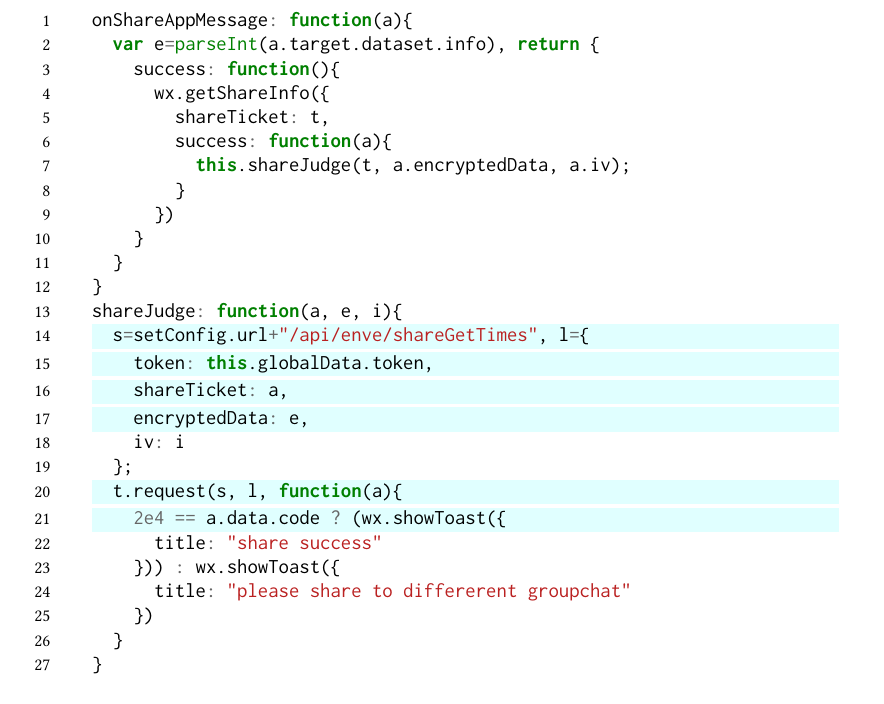} 
% \begin{minted}[linenos,fontsize=\scriptsize,highlightlines={14-17, 20-21},xleftmargin=.05\textwidth]{Javascript}
% onShareAppMessage: function(a){
%   var e=parseInt(a.target.dataset.info), return {
%     success: function(){
%       wx.getShareInfo({
%         shareTicket: t,
%         success: function(a){
%           this.shareJudge(t, a.encryptedData, a.iv);
%         }
%       })
%     }
%   }
% }
% shareJudge: function(a, e, i){
%   s=setConfig.url+"/api/enve/shareGetTimes", l={
%     token: this.globalData.token,
%     shareTicket: a,
%     encryptedData: e,
%     iv: i
%   };
%   t.request(s, l, function(a){
%     2e4 == a.data.code ? (wx.showToast({
%       title: "share success"
%     })) : wx.showToast({
%       title: "please share to differerent groupchat"
%     })
%   }
% }
% \end{minted}
 
     \caption{Promotion Abuse via \texttt{ShareInfo} Manipulation}
          \label{list:ShuiHu}
         
\end{figure}

\paragraph{Case II: Promotion Abuse via \texttt{ShareInfo} Manipulation} Aside from \texttt{WeRunData}, promotion abuse attacks can also be achieved by manipulating \texttt{ShareInfo}. We discovered that numerous mini-programs reward their users with red packets, a type of monetary gift, when they share the mini-program with their group chats. This is accomplished by fetching the group chat ID returned by \texttt{getShareInfo}. For example,  as shown in \autoref{list:ShuiHu}, we found that there is a mini-program that rewards users with red packets for sharing with a unique group chat. The mini-program's back-end verifies the encrypted data returned by \texttt{getShareInfo} to ensure that the sharing is genuine and prevent users from cheating.  However, similar to the previous case, with the leaked \appsecret, an attacker can exploit this security measure by repeatedly sharing the program and modifying the group chat ID. We have tested this mini-program by playing it normally and noticed that it gives 1 to 10 cents of random red packets to users when sharing this mini-program with a unique group. Given that it takes around 3 seconds at most to share a mini-program to a groupchat, an attacker can theoretically get \$20,000 dollars worth of red packets in a week. \looseness=-1

\subsection{Attacks against Paid Cloud Services}
\label{attack:services}

\paragraph{Attack Workflow} 
Since \appsecret is also used to access the cloud services provided by \wechat, the attacker may use the leaked \appsecret to fetch its corresponding access token \textsf{AT} and consume the paid services for free. The attacker takes place with the role of mini-program's back-end, and launches the attacks with the following two steps:

\begin{enumerate}[label=(\Roman*)]

\item \textbf{Obtaining the API Access Token (\textsf{AT}).} To obtain the  API Access Token, the attacker just needs to send the \appsecret and the victim mini-program's {\tt appID} to the \wechat server, which will reply the corresponding \textsf{AT} if both the \appsecret and {\tt appID} are legitimate, allowing attacker to gain access to all the services enabled by the victim (\textbf{Step \ding{192}}).\looseness=-1

\item \textbf{Service Invocation.}  In this stage, the attacker just needs to send a request to the \wechat server with the obtained \textsf{AT}, victim's  {\tt appID}, and the JSON object containing parameters to invoke the enabled services. %As far as the invocation quota is not exceeded, 
The server will return the result as invoked by the victim mini-program (\textbf{Step \ding{193}}).  An online OCR API charges \$1,000 for every 1 million invocations. Attackers could exploit the service by creating their own mini-program or query interface that consumes the OCR service for free using the leaked \appsecret, potentially causing a loss of thousands of dollars. \looseness=-1

\end{enumerate}

\paragraph{(A3) Service Theft Attacks} 
All vulnerable mini-programs using cloud services were found to be susceptible to service theft, allowing attackers to invoke paid services for free once they obtain the corresponding access token using the leaked \appsecret. We investigated a few examples, including Tencent's medical appointment platform, which used location services for searches, and an online game requiring identity verification using OCR (in China, an online game user may not be able to register for a game account  unless he or she presents a valid identity, which prevents children from 18 years and younger from game addiction). Additionally, several mini-game programs enabled WeChat's security service to identify user risk levels. Exploiting the leaked \appsecret, an attacker could use these services to build an online portal for commercial use without paying for anything to Tencent.

\section{Discussions}
\label{sec:discussion}

\subsection{Mitigation and Prevention}
Fundamentally, the \appsecret leakage problem is a cryptogrphic key management problem. Since the \wechat client cannot be trusted, we must rely on both the \wechat server and the mini-program back-end to properly manage the keys. At a high level, there could be three possible approaches to mitigate or even prevent \appsecret leaks.
\begin{packeditemize}
 \item {\bf Vetting the mini-programs systematically.}  One mitigation solution is to use the detection method introduced in \S\ref{Sub:methodology} to vet mini-programs. % level: 
In particular, when a mini-program is submitted, \wechat must extract the byte code from the mini-program, and detect all possible \appsecret leaks. A mini-program that fails to meet the vetting requirements will not be allowed to be released.   Since \wechat controls the life cycle of the \appsecret, it would be easier for them to perform such a detection. For example, in our analysis, to confirm whether the grepped hex strings are {\tt MK}s, we have to send each \appsecret to \wechat server, and analyze the responses. However, \wechat can fetch these keys directly from their databases. \looseness=-1

\item  {\bf Removing the encrypted data forwarding from the untrusted client when requesting sensitive data.}  According to the sensitive data access protocols described in \autoref{fig:arch}, \wechat should have redesigned Step \ding{186}-\ding{188} by removing the data forwarding step at Step \ding{187}. This is because the current design provides an opportunity for attackers to modify the ciphertext sent by the \wechat server. Consequently, \wechat should let {\tt getUserData} API communicate to the mini-program backend first, and let the back-end directly get the data from the \wechat server, instead of allowing the client to request and then forward. However, this defense may require a completely new mechanism to make it secure. Currently, \texttt{AT} is directly fetched by the MS from WS, since \texttt{AT} does not need the user's authorization (e.g., \texttt{AT} is only used to consume the services of WS). However, if we directly let the two back-ends exchange the encrypted data (without the user’s authorization, i.e., LT, sent from the front end), the MS may try to crawl all the users’ sensitive data for malicious purposes. 

\item {\bf Using signing and integrity checking.} Since the data manipulation attacks we demonstrated involve the modification of the ciphertext of the messages sent by the \wechat server, \wechat can then first sign the encrypted message using \Tencent's private key, and then ask the mini-program back-end to verify the integrity of the message with \Tencent's public key, though this can have additional computation overhead. Encouragingly, we have noticed that \wechat has released a new API \texttt{auth.checkEncryptedData}~\cite{checkEncryptedData} to check the data integrity based on this recommended defense. 
\end{packeditemize}

\begin{table}[]
\centering
    \scriptsize
    \setlength\tabcolsep{1pt}
    \aboverulesep=0pt
    \belowrulesep=0pt
\begin{tabular}{@{}llcccccc@{}}
\toprule
\textbf{Super App} & \multicolumn{1}{l}{\textbf{Name}} & \textbf{Company} & \textbf{\begin{tabular}[c]{@{}c@{}} Back-end \\ Stores MK\end{tabular}} &\textbf{\begin{tabular}[c]{@{}c@{}}Data \\   Access \end{tabular}}& \textbf{\begin{tabular}[c]{@{}c@{}}Service\\   Access \end{tabular}} & \textbf{\begin{tabular}[c]{@{}c@{}}Download\\   API \end{tabular}}     & \textbf{\begin{tabular}[c]{@{}c@{}}Verify\\   API \end{tabular}}  \\  \midrule
\textbf{WeChat~\cite{wechatkey20}}             &  \texttt{APPSECRET}   &          Tencent            & \tickYes                                                                           & \tickYes                                 & \tickYes                                    & \tickYes& \tickYes\\
\textbf{WeCom~\cite{wechatkey20}}              & \texttt{APPSECRET}  &           Tencent            & \tickYes                                                                           & \tickYes                                 & \tickYes                                    & \tickYes& \tickYes\\
\textbf{QQ~\cite{qqkey}}              & \texttt{APPSECRET} &           Tencent             & \tickYes                                                                           & \tickYes                                 & \tickYes                                    & \tickYes& \tickYes\\

\textbf{Baidu~\cite{baidusmartprogram}}              & \texttt{CLIENT\_SECRET}  &          Baidu             & \tickYes                                                                           & \tickYes                                 & \tickYes                                     & \tickYes& \tickYes\\
\textbf{Alipay~\cite{alipaykey}}             & \texttt{AES\_KEY}&             Alibaba               & \tickYes                                                                           & \tickYes                                 & \tickNo                                     & \tickYes& \tickNo\\
\textbf{Taobao~\cite{alipaykey}}             & \texttt{AES\_KEY}&             Alibaba               & \tickYes                                                                           & \tickYes                                 & \tickNo                                     & \tickYes& \tickNo\\
\textbf{DingTalk~\cite{dingdingkey}}             & \texttt{APPSECRET} &           Alibaba             & \tickYes                                                                           & \tickNo                                 & \tickYes   & \tickYes& \tickYes\\ 
\textbf{Tiktok}~\cite{bytedancekey}             & \texttt{APPSECRET}    &           Bytedance          & \tickYes                                                                           & \tickYes                                 & \tickYes   & \tickYes& \tickYes\\                             \textbf{JINRI Toutiao}~\cite{bytedancekey}               & \texttt{APPSECRET} &          Bytedance               & \tickYes                                                                           & \tickYes                                 & \tickYes   & \tickYes& \tickYes\\ 

\bottomrule
\end{tabular}
\caption{The list of super-apps that used cryptographic access control}
    \label{tab:superapps}
 
\end{table}

It is important to note that the latter two defense can only fix the attacks against sensitive data. To mitigate against attacks against paid cloud services is relatively easy, and we can adopt traditional defenses in web security. For example, in the regular case, the service request is from different IP addresses and different accounts. As such, the back-end may set up the rate limit to prevent the services from being abused. In cases where the attacker with the leaked \appsecret switches accounts and changes IP address, developers may have to update \appsecret to trade off security and usability.   
 
\begin{table}[h]
\scriptsize
 \belowrulesep=0pt
 \aboverulesep=0pt
 \setlength\tabcolsep{2pt}\resizebox{0.475\textwidth}{!}{
\begin{tabular}{llcc}\toprule
 {\textbf{API}} &  {\textbf{Description}}    \\
 \midrule

\textbf{{{M.baidu.com/sf/vsearch}}}    &   Mini-program Searching API      \\
   \textbf{Parameters}   &       \\  
\SFviii   \texttt{ word}    &   A keyword to search the mini-programs         \\  
%\SFviii Language & \texttt{obj.lang} & \tickNo    &  \tickYes     \\
%\SFviii \texttt{pn}       &   The \wechat client version    \\ 
%\SFviii \texttt{time}       & The timestamp of this query     \\ 
\SFviii  \texttt{clientip} &  \wechat client IP     \\
\SFii~ ... & Other omitted parameters\\
\textbf{Return value} &  %The  mini-program metadata  
\\
\SFviii  \texttt{appKey}     & The app ID of the mini-program       \\
\SFviii  \texttt{customer\_name}     & The developer of the mini-program       \\
\SFviii   \texttt{app\_category}     & The category of the mini-program    \\
\SFviii  \texttt{query\_click\_count}   &  The downloaded times  \\
\SFii~... & Other omitted fields in the return value\\
%\SFii Profile & A\_A1\_A\_A & a\_a1\_a   &      \\
%~~~\SFii Nick name   & A\_A1\_A\_B & a\_a1\_b   &      \\
  \midrule

\textbf{Swan.navigateToSmartProgram()}        &    Mini-program Downloading API    \\  
   \textbf{Parameters}   &       \\  
\SFviii\texttt{appKey}        &  The app ID of the mini-program          \\  
\SFii\texttt{version}    &  The version (released or debug)       \\ 
\textbf{Return value}    &  The packed file of the mini-program     \\  \midrule

 \textbf{openapi.baidu.com/oauth/2.0/token}     & A \appsecret Validation API    \\
    \textbf{Parameters}   &       \\  
\SFviii  \texttt{client\_id}      &   The app ID of the mini-program     \\
\SFviii \texttt{client\_secret}    &  A \appsecret      \\
\SFii\texttt{grant\_type}   &  The access type  \\
\textbf{Return value}    &   An access token or error message   \\

\midrule

 \textbf{Openapi.baidu.com/rest/2.0/smartapp/getsessionkey}      & \sessionkey query API (\texttt{getEK()})    \\
    \textbf{Parameters}   &       \\  
\SFviii  \texttt{appKey}      &   The app ID of the mini-program     \\
\SFviii \texttt{secret}    &  The Master key (\appsecret)      \\
\SFii\texttt{jscode}   &   A login token (\token) \\
\textbf{Return value}    &  The encryption key (\sessionkey)  \\

\bottomrule
\end{tabular} }
 
\caption{The four important APIs used in testing Baidu.}
\label{tab:inner:apis3}
 
\end{table}

\subsection{Generality of our Study}
We are confident that our findings on the \wechat mini-program paradigm can be applied to other platforms that have a similar design, such as QQ, Alipay, Baidu, Taobao, and TikTok. As shown in~\autoref{tab:superapps}, these platforms may have different names for their app secrets, but they all have sensitive data or services for their mini-programs to access and are required to store their \appsecret on their back-end systems. Thus, a compromised app secret could potentially lead to similar attacks on these mini-programs. We verified our findings by testing on  Baidu~\cite{baidusmartprogram}, which confirmed that the attacks and the methods we used to detect \appsecret leakage worked as expected. Particularly, we focused on Baidu for three reasons: (i) Baidu mini-programs are developed by Baidu Inc., which is independent from Tencent. (ii) Baidu has over 150,000 mini-programs, making it more popular than other platforms like DingTalk with only around 20,000 mini-programs. (iii) Baidu provides sensitive data and cloud services, unlike platforms such as Alipay that only offer access to sensitive data without service access. \looseness=-1

First, our attacks are applicable to Baidu as they use a similar architecture to WeChat to protect sensitive data and cloud services. Particularly, Baidu encrypts sensitive data using a user-specific encryption key, which is later decrypted at the back-end using the retrieved encryption key (the 4th API in \autoref{tab:inner:apis3}). Similar to \wechat, Baidu uses the \texttt{CLIENT\_SECRET} (i.e., \appsecret) to query access tokens to consume their services (the 3rd API in \autoref{tab:inner:apis3}). We tested our attacks on 10 Baidu mini-programs that leaked their \texttt{CLIENT\_SECRET}, and all 10 were vulnerable to our attacks (we choose not to disclose their names due to ethics concerns). For example, we have been able to execute account hijacking and service theft attacks with success. However, due to the unavailability of APIs to access \texttt{ShareInfo} and \texttt{WeRunData} on Baidu, we are unable to carry out promotion abuse attacks. The attack details are similar to those used in \wechat, so we have omitted them for the sake of brevity.   It's noteworthy that even though Baidu's documentation mentions that sensitive data is signed, they did not provide an API to verify the signature on the mini-programs' back-ends~\cite{baidusmartprogram}, which may explain why the tested mini-programs do not verify the signature and are vulnerable to attacks. \looseness=-1

Second, our methodology for collecting and verifying the correctness of the \appsecret for Baidu mini-programs is similar to that used for \wechat. Specifically, \texttt{navigateToSmartMiniprogram} API can be used to download Baidu mini-programs by providing their appIDs, making our designed crawler easily consumable for Baidu mini-programs (2nd API in \autoref{tab:inner:apis3} of Appendix-\S\ref{sec:baidu}). Additionally, Baidu also has an API to verify the correctness of the \appsecret (3rd API in \autoref{tab:inner:apis3}). This API is necessary since encryption keys or tokens cannot be delivered to untrusted mini-program front-ends. Indeed, our investigation revealed that, besides Baidu and \wechat, all other platforms have mini-program downloading APIs, and 7 out of 9 platforms have an API to verify the \appsecret. This suggests that the methodology for collecting mini-programs and verifying their secrets is similar across other platforms, in addition to \wechat and Baidu. \looseness=-1

Third, the problem of \appsecret leakage also exists in Baidu. We collected 171,989 Baidu mini-programs and used a regular expression to search for possible \appsecret{s} in the mini-program code (\texttt{[a-zA-Z0-9]{32}}), which we validated using the \appsecret validation API. Our findings show that 7,476 Baidu mini-programs  (4.35\%)  have leaked their \appsecret{s}. For readers interested in more details, we have included a detailed measurement for Baidu in our Appendix-\ref{sec:baiduleaks}.

\subsection{Ethics Concerns} 
During our experiments, we followed community practices to avoid causing harm to victims or super apps. 
\begin{packeditemize}
\item \textbf{Mini-programs Download}. we adhered to the rate limits set by the WeChat and Baidu servers, ensuring that we did not exceed the limits during our study. When downloading apps, we limited our number of requests per minute to just six while issuing. It took us more than six months to collect all the mini-programs. 
\item \textbf{\appsecret{s} Verification}. We also complied with the rate limits set by WeChat and Baidu servers for querying \texttt{AT} APIs such as \texttt{weixin.qq.com/cgi-bin/token}. When validating the \appsecret, we also limited the number of validation threads to ensure that we did not exceed the rate limit. 
For WeChat mini-programs, we reported a few mini-programs that leaked their \appsecret{s} to Tencent, and after they agreed to our approaches, we then conducted the experiment on the remaining mini-programs and reported all the mini-programs that had leaked their \appsecret{s}.
\item \textbf{Controlled Environment.}  We made sure to conduct all attacks solely using our own accounts, testing devices, and back-end servers.  We took great care to ensure that our testing activities were conducted within a controlled environment and that we did not launch any attacks against any third-party entities for the purpose of collecting or manipulating their data. \looseness=-1

\item \textbf{Responsible Disclosure.} We reported all our findings, including recommendations to fix the vulnerability, to Tencent and Baidu. They confirmed our findings and awarded us bug bounties. We were informed that they are actively fixing this problem, and mini-program developers who leaked their \appsecret{s} have been notified.
As of now, we have checked the status of some mini-programs and confirmed that some of them (such as those from Toyota, Nestle, and Cisco) have already fixed the vulnerability by removing the \appsecret{s} from the front-ends. It is worth noting that for the mini-programs that were explicitly named and not anonymized, we can confirm that they have already fixed their vulnerabilities. However, we also observed that some vendors fixed the vulnerability in the wrong way by updating the \appsecret{s} in their front-ends instead of removing them. We are also pleased to learn that WeChat has released a new API \texttt{auth.checkEncryptedData}~\cite{checkEncryptedData} to check data integrity (i.e., the third approach) based on our recommended defense. Our study has benefited the vendors greatly, and Tencent acknowledged this during our online meetings.

%We will continue to scan, report, and track this vulnerability in WeChat and Baidu platforms.
\end{packeditemize}

\looseness=-1 %Third, while we have demonstrated the feasibility of service theft attacks, we strictly limited the number of services consumption requests to one or two times. We did not use the \appsecret to build a mini-program or online services for commercial use.           \looseness=-1
%To be more specific, \Tencent state  

 %Second, we have reported all our findings to \Tencent, and we are informed that \Tencent are actively fixing this problem. \AY{To call for the awareness of mini-program security, we will release our code to Github for security researchers in the community.}\ZY{This should be claimed in the introduction. Also, the responsible disclosure section is now superficial, and I think we should add more content here.  }

%In our study, we discovered that \wechat may give developers too much freedom in implementation of security measures, and thus allowing attackers to extract keys and perform various attack against \wechat mini-programs. To mitigate the problem, \wechat should not trust developers but enforce the security measures from its own side.  Therefore, we discuss the possible solutions from the perspectives of \wechat, including the security measures that can be deployed at \wechat client (\S\ref{subsec:defenseclient}) and security measures that can be deployed at the \wechat server (\S\ref{subsec:defenseserver}).  

\ignore{
\subsubsection{\bf Mitigation at \wechat Client}
\label{subsec:defenseclient}

\wechat client maintains the mini-programs packages, and provides numerous APIs for the mini-programs to achieve different functionalities. Therefore, the front-end of \wechat can protect the keys from being leaked by using (1) code encryption and (2)  better encapsulation of APIs:

\begin{itemize}
\item \paragraph{(1) Code Encryption} One possible solution to prevent the keys from being leaked is to enforce the code encryption. This because the attacker can easily extract the leaked keys from the de-compiled byte code. Although currently, \wechat has enforced the mini-program code encryption at the \texttt{Windows} platform, the keys that are used to encrypt the source code is the mini-program IDs. These mini-program IDs are public accessible, and can even be obtained by using their built-in searching interfaces. We argue that \wechat should re-implement the code encryption mechanism, or at least select other keys instead of using the mini-program IDs directly as the encryption keys. 

%\item \paragraph{(2) Outgoing Traffic Vetting} Being the host app of all mini-programs, \wechat client can monitor all the traffics sent from the mini-programs. Although the traffic are protected by \texttt{HTTPS}, \wechat can vet the network requests from their API, since all the network request are sent from API \texttt{wx.request}. To be more specific, when the mini-program invokes \texttt{wx.request} to initiate a network request, \wechat can quickly grep the content of the traffic and search possible \appsecret leakages, since. 
%This is more powerful when compared with the code level vetting, particularly in the searching of session key leakages.  For example, in our analysis, although we can detect the potential leakage of the session keys, we have to confirm the leakage of session keys manually by intercepting the consumed interfaces. However, \wechat can directly inspect all the outgoing network requests to avoid the possible leakage of session keys.\looseness=-1

\item \paragraph{(2) Better Encapsulation of \wechat front-end API} Recall that the root cause of the attacks is that \wechat have provided too much freedom for the third party developers who are less familiar with the threat model of the mini-program. Therefore, as a solution, the developers may need to re-design the current framework to support better encapsulated APIs, where all the incoming and outgoing requests involving the key managements, data encryption and decryption are hidden from the third-party developers. Under such a framework, the \appsecret are all transparent to developers as well as attackers, and attackers need to compromise to achieve their attack goals, which is less realistic. Different from mini-programs, which are only a few MBs, and programmed using JavaScript, \wechat is a giant software (\eg, its Android version is over one hundred MB) that has numerous Java byte code, Java libraries and native C libraries. \looseness=-1   

\end{itemize}
 }

 \ignore{
\begin{table*}[h]
\small
\aboverulesep=0pt
\belowrulesep=0pt
\setlength\tabcolsep{2.5pt}
\begin{tabular}{c|l|ccccccc}
\toprule
% &permission management&user data access&peripheral device access&extended framework API&Cross-app redirection&Plugin Integration&Built-in Cloud Feature\\
\multicolumn{2}{c|}{}&mini-programs&browser extensions&Android apps&iOS apps&apps in sandbox&virtual machines\\ \midrule
\multicolumn{2}{c|}{installation source}&\wechat app&in-browser market&Google Play&App Store&---&---\\ \midrule
\multirow{3}{*}{host app}&permission management&\CIRCLE&\Circle&\CIRCLE&\CIRCLE&\Circle&\Circle\\
&sensitive user data access $^{*}$&\CIRCLE&\Circle&\CIRCLE&\CIRCLE&\Circle&\Circle\\
&peripheral device access&\CIRCLE&\Circle&\CIRCLE&\CIRCLE&\LEFTcircle&\LEFTcircle\\ \midrule
\multirow{2}{*}{front-end}&extended functional API&\CIRCLE&\LEFTcircle&\CIRCLE&\CIRCLE&\Circle&\Circle\\
&cross redirection&\CIRCLE&\CIRCLE&\CIRCLE&\CIRCLE&\LEFTcircle&\Circle\\ \midrule
\multirow{2}{*}{back-end}&built-in cloud feature&\CIRCLE&\Circle&\Circle&\Circle&\Circle&\Circle\\
&oficial network API domain&\CIRCLE&\Circle&\Circle&\Circle&\Circle&\Circle\\
\bottomrule
\end{tabular}
\caption{Comparison of features provided by each paradigm framework}
\label{tab:comparison}
\end{table*}
  }

\ignore{

\subsubsection{Mitigation at \wechat Server}
\label{subsec:defenseserver}

%From the \S\ref{sec:back}, we know that \wechat Server plays a vital role in the ecosystem of mini-programs. First, all the mini-programs that submitted from different developers need to be reviewed by the \wechat server, and only these quantified with lower risks can be published. Second, \wechat server also needs to communicate both front-end and back-end, particularly the back-end, to provide the third party developers the users sensitive data. Therefore, our possible mitigation also covers these two aspects: 

\paragraph{Mitigation for mini-program Front-end Vetting} \wechat server reviews all the mini-programs that are submitted from the developers. However, we argue that the current vetting process of \wechat is not good enough, which have ignored many key leakage cases. We believe the vetting process can be improved from both the code level and the IDE level:

\begin{itemize}
\item  \paragraph{(1) Code Level Vetting }  One solution is to use the detection method introduced in our paper to conduct a vetting process over the code level: when third-party developers submit the mini-program, \wechat extract the byte code from the mini-program, and detects all the possible \appsecret leakages.  mini-programs that fail to meet the vetting requirements will not be allowed to be released.   Since \wechat controls the life cycle of the   \appsecret, it would be easy for them to enable such detection. For example, in our analysis, to confirm whether the grepped strings are \appsecret, we have to send the \appsecret to \wechat interface, and analyze the responses. However, \wechat can fetch these keys directly from their databases. 

%\item \paragraph{(2) IDE Level Vetting} To prevent developers from   \appsecret leakage to front-end, \wechat can also enforce the detection of key leakage in their programming IDE and SDK. For example, the leakage of session key and \appsecret can be mapped to programming grammar errors, and mini-programs with this key of errors cannot pass the compiling process. Although we are aware of the confirmation of the session keys and \appsecret requiring the communication between IDE and \wechat server, which may downgrade performance of IDE, we argue that \wechat can at least prompt warnings when the IDE detects the code fragments that may have potential leakages. For example, the offline IDE can also easily grep the 32-hex strings (recall that all \appsecret have the same format) existing in the source code of the mini-program, and deliver warnings to remind developers of the leakage of \appsecret.   

\end{itemize}

\paragraph{Mitigation for Protecting mini-program Back-end} \wechat server handles the communication between the front-ends and back-ends of all the mini-programs. As the front-ends of mini-programs are not trusted, the \wechat server can either weaken the involvement of front-ends, or strengthen and enforce the security measures at the back-ends of the mini-programs. 

\begin{itemize} 
\item 
\paragraph{(1) Weakening the Involvement of Un-Trust Parties} One of the intuitive solutions is to avoid the unnecessary involvement of the front-end. Currently, the un-trust party, the front-end of mini-program, is still heavily involved during the data encryption and decryption: the login token and the cipher-text of the sensitive data are all sent from the front-end. This gives the chance for attackers to decrypt the cipher-text and manipulate the sensitive data. Although the token cannot be removed from the framework design, as the token is used to associate the user's identifier, the cipher-text transportation can be removed from the front-end. That is, when the mini-program attempts to access the sensitive data, the front-end first send the request to its back-end. The back-end receives the request, and  then fetches the cipher-text directly from the back-end, which gives the attacker no chance to manipulate the sensitive data from the front-end.  

\item  \paragraph{(2) Public Key Infrastructure } 
Another solution to prevent \appsecret from being leaked is to use the Public Key Infrastructure (PKI).  Recall that \appsecret is used to authenticate the developers. However, we believe such a authentication mechanism can be replaced with the PKI. Instead of making \appsecret as a constant string and requiring the developers to maintain it,  the private key can be installed onto the developer's machine that runs the back-end of mini-program. \wechat can easily integrate the off-the-shelf open SSL implementation, and authenticate the developers easily without even involving the interactions from the developers.  

\item  \paragraph{(3) Integrity Checking} \wechat can also enforce a built-in integrity checking mechanism for the third-party developers to prevent the transferred encrypted messages from being modified by the attackers.   
 Particularly, we emphasis that such an integrity checking process has to be a built-in one; Otherwise, the developers may misuse the mechanism too. For example, \wechat can first sign the cipher,  then deliver the singed cipher to the front-end, and ultimately verify the cipher at the \wechat gateway. Any modified cipher-text will not pass the integrity checking and be rejected by \wechat.  In this case, the private key used to generate the signature is maintained by \wechat server, which will never be leaked to any front-ends.

\end{itemize}

%\ZY{I do not think this part is needed. First, it is not related to our paper. Second, Luyi also talked the similar content in their background. At that time, their reviewers may challenged the differences between the mini-programs and existing paradigm, so they added that part. But, currently, since his paper has already been published, I do not think this part is still needed. I think we can put this part at the discussion section.}\ZQ{maybe we can keep it, but the description has to be crystal clear, and also serve the purpose for this paper, particularly if we have some background knowledge to refer here. If no, then we can remove it}

%\AY{I added this section just to describe that mini-programs have complex functionalities of both native app functionalities and unified cloud function.}
%\subsection{Encryption Key (\sessionkey) Leakage} 

%In our study, we have demonstrated that \appsecret can be leaked and be used to mount various attacks. However, other than \appsecret, we also found that \sessionkey can be misused. Recall that \sessionkey is used to encrypt the sensitive data transferred between the mini-program front-end and its back-end.  Although \sessionkey is not fixed and cannot be hard-coded in the mini-program's front-end, we interestingly found some developers may mistakenly implement a interface that is used to query the \sessionkey from the backend,  resulting in the leakage of the session key.  

\subsection{mini-program vs Existing Paradigms} 

Similarly to the mini-program paradigm, there had been other approaches to allow applications running in a execution environment in a certain framework. Aside from classical native apps that run in OSes such as Android and iOS, there have been two other major paradigms that are similar to mini-program paradigm: the browser extension and the virtual machine. Virtual machines improve security by allowing Operating Systems to execute in a isolated software-simulated PC or mobile device, whereas ensuring that the virtual machines are unaware of the framework. Hence, this paradigm is widely used in security vulnerability analysis and education. On the other hand, the browser extension framework extends the functionality by allowing extensions to execute on top of the web browser. As such, these browser extensions can provide many functionalities that pure webpages cannot provide. For example, with browser extensions, users can schedule meeting, view movies, and even play games \cite{}.

As shown in \cref{tab:comparison}, we compared the mini-program paradigm with other five similar paradigms in terms of supported features provided for mini-programs or containers running within the frameworks. As we can see, different from sandboxes and virtual machines, \wechat mini-program extends the functionalities with the mini-program APIs, significantly expanding the functionalities of mini-programs. Compared with mobile native apps such as Android and iOS apps, mini-program framework shares a lot of similarities in that it also integrates a set features such as permission management, sensitive user data access (e.g., user phone number), and cross-app redirection. However, different from the mobile native apps that merely provide basic network request APIs, mini-programs provides built-in cloud feature and a unified network API domain that provides various network APIs for third-party back-end servers. This enables \wechat to secure sensitive user information by implementing additional security mechanisms such as data encryption and invoker verification.
}

 %\vspace{-0.1in}   
 \section{Related Work}
  %  \vspace{2mm}
    \label{sec:related}

\paragraph{Mini-Program Studies}  
Lu et al.~\cite{lu2020demystifying} examined the security of the mini-program paradigm by focusing on access control mechanisms.  More recently, Zhang et al.\cite{Identity:confusion:2022} explored identity confusion in WebView-based super apps, while Yang et al.~\cite{yang2022cross} discovered a new cross mini-app request forgery attack. Other efforts have been made to understand this novel paradigm. For instance, Zhang et al.\cite{ourfancywork} developed {\sf MiniCrawler} to download mini-programs and conducted a large-scale measurement, while Hao et al.\cite{hao2018analysis} studied the system architecture and key technologies used by \wechat mini-programs.  Liu et al.\cite{liu2020industry} created a dynamic analysis framework for \wechat mini-programs and evaluated their tool on 152 open-source mini-programs. 
Additionally, some studies aim to understand specific applications and use-cases, including healthcare~\cite{wang2017evaluation,zhou2020benefits}, transportation~\cite{cheng2019exploratory}, marketing~\cite{rao2021impulsive}, and education~\cite{liang2019construction,sui2020impact,chen2020value}. Our work investigates the misuse of cryptographic keys, specifically the \appsecret, by mini-programs and their potential attack consequences.

\paragraph{API Misuse} API misuse is a critical topic in software development, as highlighted in several studies~\cite{sven2019investigating,zhang2018code,li2021arbitrar,bae2014safewapi,ren2020api,gorski2018developers}. Such misuse can result in security vulnerabilities, as evidenced by the discovery of API misuse in third-party mobile payment systems that allowed attackers to bypass payment processes~\cite{yang2017show}, and the exploitation of Android fingerprint APIs to launch attacks on vulnerable apps~\cite{bianchi2018broken}. Researchers have developed various tools and techniques to detect and prevent API misuse. For instance, \texttt{SAFEWAPI}\cite{bae2014safewapi} uses Web API specifications to identify API misuses in JavaScript web applications, while \textsf{ARBITRAR}\cite{li2021arbitrar} leverages active learning to classify valid and invalid usages of target API methods. Ren et al.\cite{ren2020api} construct a knowledge graph from API reference documentation to enhance API misuse detection, while \textsf{MuDetect}\cite{sven2019investigating} employs a graph-based representation and ranking strategy to detect API misuses. Additionally, Gorski et al.~\cite{gorski2018developers} propose an API-integrated security advice approach to help software developers write more secure code for difficult-to-use cryptographic APIs.  \textsf{ExampleCheck}~\cite{zhang2018code} analyzed over 217,818 Stack Overflow posts and discovered that 31\% of them potentially contained API usage violations.

\paragraph{Credentials Leakage and Detection} Credentials may be hardcoded in open-source projects, which can lead to security vulnerabilities. Sinha et al. ~\cite{sinha2015detecting} and Singh et al.\cite{singh2021revisiting} discuss the issue of API key leaks, in which malicious users steal API keys from public code repositories. Shi et al.~\cite{shi2021empirical} studied leaked payment credentials for Cashiers serving over one billion users using \texttt{PayKeyMiner}. Wen et al.~\cite{wen2022secrethunter} developed a real-time, large-scale comprehensive secret scanner for GitHub called \texttt{SecretHunter}.
Michael et al.~\cite{ndssleak} detected potential credential leakage (such as SSH keys and API tokens) from GitHub repositories with a hybrid approach that involves both automatic detection and manual validation. Basak et al.~\cite{basak2022practices} studied current key management practices by analyzing key leakage in GitHub. Lounici et al.~\cite{lounici2021optimizing} and Saha et al.~\cite{saha2020secrets} detect key leakage in GitHub using machine learning. \texttt{PassFinder}~\cite{feng2022automated} employs deep neural networks to effectively detect user password leakage from public repositories on GitHub. Compared to previous studies that focused on key leakages in Github, our study is unique in that we focus on mini-programs instead. Also, unlike other studies that aim to improve detection methods, we use simple methods to detect keys and ensure no false positives by validating the correctness of keys using API.  

Credentials may also be hardcoded in the front-end of mobile apps, which can lead to security vulnerabilities. Viennot et al.~\cite{viennot2014measurement} performed the first study of secret authentication key usage and its problems in Android apps using \texttt{PlayDrone}. Zhou et al.~\cite{zhou2015harvesting} developed a tool called \texttt{CredMiner}, which can programmatically identify and recover developer credentials from native Android apps. They identified 302 email credentials and 58 Amazon AWS credentials over 36,561 native apps. Zuo et al.~\cite{zuo2019does} conducted a study on cloud service credentials in mobile apps and discussed their root causes and impacts. Wen et al.~\cite{wen2018empirical} scanned 100 popular iOS apps and identified that 48 of them had misused credentials, including credential leakage, using \texttt{iCredFinder}. Nan et al.~\cite{nan2018finding} developed a semantics-driven solution that utilizes NLP to automatically discover credentials in modern mobile apps. Moreover, Wang et al.~\cite{wangcredit2023} recover cloud credentials from apps, infer their capabilities in the cloud, and verify if the capabilities exceed the legitimate needs of the apps. Once again, the focus of other works is primarily on detection methods, such as using machine learning or NLP. In contrast, our focus is on the problem space, where we demonstrate how credentials can lead to various attacks in mini-programs.

%These credentials are often third-party credentials, making it difficult for mobile app providers (e.g., Google Play) to detect their validity and assess their impact. However, in our study, the leaked credentials are first-party credentials, and it should not be difficult for super apps to identify them. This problem has persisted since 2017 based on our measurements, and developers have never fully understood it and continue to make the same mistake. We believe that the main reason for this is that they have not been educated about the possible outcomes of the problem. We hope that our work can raise awareness among developers and help them avoid such issues in the future.

%As presented in \autoref{tab:compare}, o
To summarize, our work is distinct from previous efforts for three reasons. First, unlike previous leakages that were detected in Github or mobile app code, our \appsecret leaks occur in the novel mini-program paradigm. Second, the focus of our study differs from most other works. They mainly focus on improving credential detection methods. Instead, we aim to understand the development practices and unique protocols involving credentials in mini-programs, as well as the security implications of leaked credentials.
Third, the protocols we analyzed are significantly different from previous studies.   As discussed in \S\ref{sec:back}, mini-programs have inherent support for key management protocols, encryption-based resource access protocols, and token-based resource access protocols that are not available in mobile apps or web browsers. In previous efforts, leaked credentials from third-party services such as Amazon or Facebook were detected. Those protocols  are also novel. Understanding those protocols will help us to comprehend the security issues and threats that the novel mini-program paradigm may face.   
Fourth, our attacks have different security consequences. While in previous efforts, attackers may use leaked keys to consume services, which is similar to our service theft attack, 1st parties such as Android or iOS do not distribute keys to their developers to access resources. Therefore, our attacks are novel because they take advantage of new features in the novel mini-program paradigm (e.g., in a promotion abuse attack, vendors use the novel werun data to attract users).  

\section{Conclusion}
  %  \vspace{2mm}
    \label{sec:conclusion}
    We have presented the first systematic study of the cryptographic access control in mini-programs, and found that \wechat has made the developers heavily involved in the security related implementation including key management, encryption, and decryption. As such, the developers can mistakenly leak their master keys to the untrusted front-end, leading to various attacks such as %which allows attackers to manipulate the sensitive resources for fun and profit such as 
account hijacking and promotion abuse. %, or consume paid services for free. 
%
%We present an in-depth measurement study of the root cause of the appsecret leakage attack, the prevalence among mini-program developers, and the consequences. %
%\AY{By utilizing the network APIs from \wechat, we crawled and analyzed over 2 million mini-programs on the mini-program market.} \sout{By providing the keywords using the company name of Fortune-500, we crawled and analyzed \totalanalyzedapps mini-programs.} 
With a large-scale measurement study, we have discovered that 40,880 mini-programs % (including many high-profile mini-programs such as tTencent and Nestle) 
leaked their {\appsecret}s. %, and perform various attacks against those mini-programs. 
%
%
%We have disclosed these \appsecret leakage vulnerabilities and also the corresponding vulnerable mini-programs to Tencent, and awarded with the bug bounty. We also learn that Tencent is actively working with the developers to fix this vulnerability and some of them have already fixed it. \looseness=-1 % \AY{With our disclosure being confirmed by \Tencent,} the corresponding fix is \sout{still} in progress.
%Our results are worrisome: we found that even high profile companies including \texttt{Nestle}, \texttt{HP} made the mistakes. 
In addition to the responsible disclosure of this vulnerability and also the list of the vulnerable mini-programs to \Tencent, we have also discussed the possible countermeasures, particularly on how \Tencent could have fixed this issue. So far, \Tencent has provided a new API to mitigate the attacks based on our recommendations. \looseness=-1

%We have also analyzed the root cause of this vulnerability and discussed possible countermeasures. Finally, % approaches to mitigate this problem. 
 %we have also made the responsible disclosure to \Tencent and awarded with the bug bounty.      

 %\input{paper/04-crawler}

%\input{ccs-body.tex}
%\bibliographystyle{IEEEtranS}
% \bibliographystyle{ACM-Reference-Format}
\bibliographystyle{plain}
%\hfill
{
\bibliography{paper, firmware, ble}
}

\appendix

\section{Measuring The \appsecret Leaks in Baidu}
\label{sec:baiduleaks}

In a similar manner to our measurement of \appsecret leaks in \wechat, we also measured \appsecret leaks in Baidu. However, as Baidu currently does not have the last update timestamp in its metadata, we conducted a measurement study to answer just the following five research questions:

\begin{itemize}
\item \textbf{RQ1:} What are the categories of the \appsecret leaked mini-programs?
\item \textbf{RQ2:} What are their popularity?
\item \textbf{RQ3:} What are their accessed resources?
\item \textbf{RQ4:} Who are their developers?
%\item \textbf{RQ5:} When are their latest update?
\item \textbf{RQ5:} Are there any high profile \appsecret leaked mini-programs?

\end{itemize}
 
\begin{table}[h]
    \centering
    \footnotesize
    \begin{tabular}{lrrrr}
\toprule
\textbf{Category} & \textbf{UserInfo} (\#) &  \% & \textbf{PhoneNumber} (\#) &   \% \\
\midrule
Automobile & 81 & 1.42\% & 40 & 1.01\% \\
Business & 620 & 10.84\% & 398 & 10.01\% \\
Charity & 1 & 0.02\% & 1 & 0.03\% \\
E-commerce & 26 & 0.45\% & 15 & 0.38\% \\
Education & 261 & 4.56\% & 168 & 4.22\% \\
Efficiency & 175 & 3.06\% & 92 & 2.31\% \\
Entertainment & 62 & 1.08\% & 33 & 0.83\% \\
Finance & 5 & 0.09\% & 2 & 0.05\% \\
Food & 44 & 0.77\% & 24 & 0.60\% \\
Government & 45 & 0.79\% & 29 & 0.73\% \\
Health & 1 & 0.02\% & 0 & 0.00\% \\
Information & 289 & 5.05\% & 114 & 2.87\% \\
IT tech & 21 & 0.37\% & 12 & 0.30\% \\
Lifestyle & 367 & 6.42\% & 253 & 6.36\% \\
Medical & 47 & 0.82\% & 5 & 0.13\% \\
News & 2 & 0.03\% & 2 & 0.05\% \\
Post service & 30 & 0.52\% & 23 & 0.58\% \\
Real estate & 340 & 5.95\% & 440 & 11.06\% \\
Shopping & 1,793 & 31.35\% & 1,222 & 30.72\% \\
Social & 53 & 0.93\% & 38 & 0.96\% \\
Sports & 25 & 0.44\% & 19 & 0.48\% \\
Tool & 11 & 0.19\% & 2 & 0.05\% \\
Traffic & 29 & 0.51\% & 22 & 0.55\% \\
Travelling & 636 & 11.12\% & 607 & 15.26\% \\
Uncategorized & 755 & 13.20\% & 417 & 10.48\% \\

\hline
Total & 5,724 & 100\% & 3,996 & 100\% \\
\bottomrule
\end{tabular}
    \caption{Statistics of the identified vulnerable Baidu mini-programs w.r.t their categories. The percentage represents the proportion of vulnerable mini-programs accessing a particular resource, compared to the total number of mini-programs accessing that resource.}
    \label{tab:baiducategory}
\end{table}

\noindent In total, we found that 7,476 of the 171,989 Baidu mini-programs (4.35\%) we collected have leaked their \appsecret{s}. % Our next step is to address those research questions by examining the categories, popularity, accessed resources, and developers of the mini-programs.
In the following, we provide the answers for these five RQs with respect to these mini-programs. % a detailed analysis of these results and answer the five RQs.

\paragraph{(RQ1) Categories of the \appsecret Leaked Mini-programs}  Similar to \wechat, most Baidu mini-programs have been classified into specific categories, as shown in \autoref{tab:baiducategory}. Since Baidu does not permit individual developers to release their mini-programs, all of them are developed by different companies. It is noticeable that shopping, business, and traveling mini-programs have a slightly higher number of vulnerable mini-programs due to the larger number of mini-programs in those categories. As these mini-programs typically involve payments, attacking them through their leaked \appsecret may have broad security implications.

\begin{table}[]
    \centering
    \footnotesize
 
 \begin{tabular}{lrrrr}

\toprule
\textbf{Download Times}   &   \textbf{ UserInfo (\#)} &    \textbf{\%}   & \textbf{PhoneNumber (\#)} &   \textbf{\%}   \\
\midrule
 (0, 1,000]           &            1,479 &     25.86\%            &          1,002 &  25.19\%               \\
  (1,000, 2,000]        &             952 &           16.65\%            &     647 &  16.26\%               \\
  (2,000, 3,000]        &             785 &             13.73\%            &   522 &  13.12\%               \\
  (3,000, 4,000]        &             430 &           7.52\%             &     319 &  8.02\%                \\
 (4,000, 5,000]        &             545 &            9.53\%             &     408 & 10.26\%               \\

 (5,000, 10,000]       &             867 &          15.16\%            &      634 &  15.94\%               \\

 (10,000, 100,000]     &             644 &          11.26\%            &       436 & 10.96\%               \\
 (100,000, 1M]   &              16 &          0.28\%             &         9 & 0.23\%                \\
 > 1M      &               1 &            0.02\%             &      1 &  0.03\%                \\
 \midrule
 {Total} & 5,719 & 100\% & 3,578 & 100\% \\
\bottomrule
\end{tabular}
    \caption{Statistics of the identified vulnerable Baidu mini-programs w.r.t their download times. Please be aware that a few mini-programs do not have their download times available, which is why the total number of mini-programs accessing user info and phone number is less than the number presented in \autoref{tab:baiducategory}.  }
    \label{tab:baidupopularity}
\end{table}

\paragraph{(RQ2) Popularity of the \appsecret Leaked Mini-programs} As illustrated in \autoref{tab:baidupopularity}, the statistics of identified vulnerable Baidu mini-programs with respect to their download times, which signify their popularity, are reported. It is evident that the majority of vulnerable Baidu mini-programs exhibit relatively low download times, with over 50\% of them residing within the (0, 1,000] and (1,000, 2,000] ranges. This finding implies that less popular mini-programs are more susceptible to security vulnerabilities. Nevertheless, it is crucial to emphasize that even in the (100,000, 1M] and >1M ranges, instances still exist. This observation serves as a reminder that security issues can persist, regardless of the high download rates of certain mini-programs.

\begin{table}[]
\footnotesize
 \setlength\tabcolsep{2pt}
\begin{tabular}{@{}llc@{}}
\toprule
\textbf{Service Name} & \textbf{Description}                                                                            & \begin{tabular}[c]{@{}c@{}}\textbf{Enabled} \\\textbf{by default}\end{tabular} \\ \midrule
Login                 & Services for users to Login                                                                     & \tickYes     \\
Book Shelf            & \begin{tabular}[c]{@{}l@{}}Online bookshelf services \\ for e-learning apps\end{tabular}        & \tickNo      \\
Coupon                & \begin{tabular}[c]{@{}l@{}}Services for developers manage coupons\end{tabular}          & \tickYes     \\
Resource Management   & \begin{tabular}[c]{@{}l@{}}Services for developers manage online resources\end{tabular} & \tickYes     \\
Template Messages     & \begin{tabular}[c]{@{}l@{}}Services for manage template  messages\end{tabular}                & \tickYes     \\
Customer Messages    & \begin{tabular}[c]{@{}l@{}}Services for manage customer messages\end{tabular}               & \tickYes     \\
Risk Detection        & \begin{tabular}[c]{@{}l@{}}Services for detecting malicious users\end{tabular}            & \tickYes     \\
QR Code               & Services for creating QR code                                                                   & \tickYes     \\
Comment Management    & \begin{tabular}[c]{@{}l@{}}Services for managing comments \end{tabular}  & \tickYes     \\
Content Security      & Services for detecting illegal content                                                          & \tickNo      \\ \bottomrule
\end{tabular}
\caption{Summary of the Baidu-managed cloud services.}
\label{tab:baiduservices}
\end{table}

\begin{table}[] 
    \footnotesize
\begin{tabular}{rrrr}

\toprule
  \textbf{ Number of Apps} &   \textbf{Downloaded \ensuremath{<}= 5000} &   \textbf{Downloaded \ensuremath{>} 5000} &   \textbf{Total} \\
\midrule
                1 &                 3981 &                1422 &   5403 \\
                2 &                  183 &                 126 &    309 \\
                3 &                  55  &                  35 &    90 \\
                4 &                  27  &                  19 &     46 \\
                5 &                  23  &                  21 &     44 \\
                6 &                  20  &                  7 &     27 \\
                7 &                   9  &                   7 &     16 \\
                8 &                  10  &                   7 &     17 \\
                9 &                   6  &                   5 &     11 \\
               10 &                   5  &                   4 &      9 \\
            $> 10$ &                  20  &                  10 &     30 \\
\bottomrule
\end{tabular}
    \caption{Numbers of vulnerable mini-programs associated with the same software development companies. Note that not all apps have their developer information available.}
    \label{tab:baidudevelopers}
\end{table}

\paragraph{(RQ3) Accessed Resources of the \appsecret Leaked Mini-programs} %The sensitive resources include sensitive data, which need to be consumed by an encryption key, and cloud services, which need to be consumed by an access token. 
Both \autoref{tab:baiducategory} and \autoref{tab:baidupopularity} display the accessed resources of vulnerable Baidu mini-programs. It can be observed that they access more \texttt{UserInfo} than \texttt{PhoneNumber}. 
The discrepancy could suggest that developers exercise greater caution when handling sensitive data, such as phone numbers. Nevertheless, a significant number of mini-programs still access this information. Meanwhile, we would like to note that since Baidu lacks APIs for accessing \texttt{ShareInfo} and \texttt{WeRunData}, we only provide statistics on the number of mini-programs that utilize \texttt{UserInfo} and \texttt{PhoneNumber}.  In contrast to WeChat, Baidu does not offer APIs such as  \texttt{api.weixin.qq.com/cgi-bin/get\_current\_selfmenu\_info} to \\verify the enabled services of specific mini-programs. The sole method to determine the availability of services is to invoke them, which can raise ethical concerns. Consequently, we cannot showcase the enabled services of these mini-programs. 

However, as demonstrated in \autoref{tab:baiduservices}, we notice that numerous services are enabled by default, indicating that attackers can launch service theft with success. %We now explore their security implications.
It is evident that some of these services are employed to manage essential online resources (e.g., images, videos) within the mini-programs. An attacker could potentially delete these resources by exploiting service theft attacks. Additionally, certain APIs can send messages (e.g., template message services and customer message services) directly to mini-programs running on users' devices, potentially tricking users into believing and clicking on phishing websites (i.e., the attacker may masquerade as an official party to send phishing messages). Interestingly, many of these APIs  (e.g., resource management services) have a rate limit, meaning that if an attacker actively utilizes these services, developers may lose the opportunity to invoke them.  

\paragraph{(RQ4) The Developers of the \appsecret Leaked Mini-programs} As shown in \autoref{tab:baidudevelopers}, the data is categorized into the number of vulnerable mini-programs and the number of companies, which are separated into two groups: those with mini-programs downloaded less than or equal to 5,000 times, and those with mini-programs downloaded more than 5,000 times. Fewer companies have multiple vulnerable mini-programs. For instance, there are 309 companies with 2 vulnerable mini-programs, 110 companies with 3 vulnerable mini-programs. This trend could suggest that having multiple vulnerable mini-programs is relatively rare.  There are also 30 companies with more than 10 vulnerable mini-programs. For example, a company named ``ShangFang (ShenZhen) Tech Inc.'' developed 87 vulnerable mini-programs. 

 \paragraph{(RQ5) High Profile \appsecret Leaked Mini-programs} We also discovered that 21 mini-programs were developed by Fortune 500 companies, including many prominent names such as Sony, HP, Amazon, and Toyota. These findings indicate that even large, well-established companies can make mistakes when it comes to developing secure mini-programs. Among the 21 mini-programs, 16 of them accessed the user's phone number or personal information. In addition to these well-known companies, Baidu, the provider, developed 39 vulnerable mini-programs. Of these, 33 accessed the user's phone number or personal information. This highlights the importance of addressing security concerns across organizations of all sizes and reputations in order to protect user privacy and maintain trust.

\end{document}